\documentclass[aps,amsmath,pra,nofootinbib,reprint,longbibliography]{revtex4-2}
\usepackage{graphicx}% Include figure files
\usepackage{dcolumn}% Align table columns on decimal point
\usepackage{bm}% bold math
\usepackage{verbatim}
\usepackage{dsfont}
\usepackage{wrapfig}
\usepackage{lipsum}
\usepackage{natbib,bbold}
\usepackage{lipsum}
\RequirePackage{float}
\usepackage[colorlinks=true,urlcolor=blue,linkcolor=red,citecolor=blue]{hyperref}
\begin{document}
\title{Quantum coherence measures for generalized Gaussian wave packets under a Lorentz boost}% Force line breaks with \\

\author{Arnab Mukherjee}
\email{mukherji.arn@gmail.com}

\author{Soham Sen}
\email{sensohomhary@gmail.com,soham.sen@bose.res.in}

\author{Sunandan Gangopadhyay}
\email{sunandan.gangopadhyay@gmail.com}

%\author{Archan S Majumdar}
%\email{archan@bose.res.in}

%\author{Sohom Sen}
 
\affiliation{
Department of Astrophysics and High Energy Physics\\
S. N. Bose National Centre for Basic Sciences, JD Block, Sector-III, Salt Lake, Kolkata 700106, India} %and/or address\\
\begin{abstract}
\noindent In this paper we consider a single particle, spin-momentum entangled state and measure the effect of relativistic boost on quantum coherence. The effect of  the relativistic boost on single-particle generalized Gaussian wave packets is studied. The coherence of the wave function as measured by the boosted observer is studied as a function of the momentum and the boost parameter. Using various formulations of coherence, it is shown that in general the coherence decays with the increase in momentum of the state, as well as the boost applied to it. A more prominent loss of coherence due to relativistic boost is observed for a single particle electron than that of a neutron. The analysis is carried out with generalized Gaussian wave packet of the form $\mathcal{N} p^n \exp(-\frac{p^2}{\sigma^2})$ with $n$ being the ``generalization parameter" and $\mathcal{N}$ denoting the appropriate normalization constant. We also obtain a range for parameter $n$ appearing in the wave packet. The upper bound is found to have a dependence on the mass of the particle and the width of the Gaussian wave packet. We have obtained the Frobenius-norm measure of coherence, $l_1$ and $l_2$ norm measure of coherence, and relative entropy of coherence for a (1+1) and (3+1)-dimensional analysis. Corresponding to each of the cases, we observe that the $l_1$ norm measure of coherence is equal to the Frobenius norm measure of coherence. We have analyzed the scenario for which such a beautiful coincidence can occur. Finally, we have plotted different measures of coherence for the electron as well as the neutron for different values of the width of the wave-function $\sigma$, boost parameter $\beta$, and generalization parameter $n$.
\end{abstract}

\maketitle

%\tableofcontents

\section{Introduction}
\noindent A very few fundamental questions raised in the previous century \cite{AEinsteinPR,ESch,Bell,Clauser}  are considered to be the origin of quantum information science. Over the last few decades, this particular branch of science has gone through an unprecedented amount of development among which the majority of the research considers a nonrelativistic background. 
However, we know that the special theory of relativity and quantum mechanics both hold up to unbelievable accuracy in real-world observations. On one hand, special relativistic analysis becomes crucial at relative velocities (between inertial frames) to be of the order of the speed of light and on the other hand, non-relativistic quantum mechanics considers physical phenomena at velocities much smaller than the speed of the light. Interestingly quantum entanglement, built up based on non-relativistic considerations, do have dependence on the reference frame of the observer. It was observed in \cite{GingrichAdami,Peres} that the spin of a massive particle depends on the momentum change caused by Lorentz transformation. This occurs due to the dependence of the Lorentz transformation of the spin of a particle on its momentum \cite{MCzachor,Peres2,Alsing,Dahn,Alsing2,Peres3,Pachos,Gonera,Lamata,Bartlett,TFJordan,TFJordan2}. In \cite{MCzachor}, the relativistic generalization of the Einstein-Podolsky-Rosen (EPR) experiment was first considered. Later in \cite{Fuentes,Fuentes2,Fuentes3}, the observer dependence of entanglement was considered for both Klein-Gordon and Dirac particles (for observers near the event horizon of a black hole). This relativistic description of quantum information theory also helps in enhancing the black hole information paradox \cite{Lloyd,ECP} and quantum gravity phenomenology \cite{Livine}, and cosmology \cite{ASM,Maldacena} as well. One of the most unique property that a quantum system may carry is quantum coherence which is, by default, not present in a classical system. It has been observed in \cite{TBaumgratz} that quantum correlations arise in composite systems due to quantum coherence. One can also exploit this quantum coherence to complete complicated quantum tasks as well \cite{Mhorodecki,Brandao,Gour,Gour2,Spekkens,Spekkens2,Spekkens3,Marvian,Marvian2}.  Although there are no unique description of coherence, there have been several basis dependent \cite{BD} and independent \cite{BID,BID2} measures of coherence. In \cite{RiddhiASM}, we observe an endeavour to devise a relativistic description of this quantum coherence for single particle, entangled states. In \cite{RiddhiASM}, the authors analyzed the change in coherence of a single-particle Gaussian states when Lorentz boost is applied. A generic loss of coherence was observed for a relativistic observer. In this work, our main motivation is to extend this analysis of relativistic coherence to much generalized Gaussian-like wave packets. We have considered here generic Gaussian wave packet structure (centred at the origin) where the exponential function is  being multiplied by $p^n$ (for general $n$ values). The main motivation behind this consideration are the Fourier transformed form of the Hermite polynomials which have similar structure to that of the wave function being considered in this work. We also have explored the effect of the ``generalization parameter" $n$ on the value of the coherence measure. Our analysis revolves around a basis independent analysis. The authors in \cite{RiddhiASM} has considered a state for which the spin state is ($|0\rangle$) for a (3+1)-dimensional scenario. In our case, we have considered a state where the spin part has maximum coherence compared to the case discussed in \cite{RiddhiASM}. To investigate the true dependence of the coherence measure for a generalized wave-function, we have calculated different coherence measures. Primarily, we have calculated the Frobenius-norm and $l_1$ norm measure of coherence where the maximum value of a coherence-measure is unity. We have also calculated the $l_2$ and relative entropy coherence for the (1+1) as well as (3+1)-dimensional cases.
The Frobenius-norm measure of coherence is defined as \cite{BID}
\begin{equation}\label{1.1}
C_F(\rho)\equiv\sqrt{\frac{d}{d-1}}||\rho-\rho_*||_F
\end{equation}
where $d$ denotes the dimension of the Hilbert space and $\rho_*=\frac{\mathbb{1}_d}{d}$ denotes the maximally mixed state. The Frobenius norm is defined as 
\begin{equation}\label{1.2}
||\mathcal{B}||_F=\sqrt{\text{tr}\left[\mathcal{B}^\dagger\mathcal{B}\right]}~.
\end{equation}
 The Frobenius-norm measure of coherence is basis independent and is invariant under unitary transformation ($C_F(U\rho U^\dagger)=C_F(\rho)$). The maximally mixed state is considered to be fully incoherent state. Hence, the coherence measure directly takes into account the geometric distance (via the Frobenius norm) between the state $\rho$ from the completely incoherent state $\rho_*$.  Hence, not only the Frobenius norm measure of coherence has an analytical expression such that $C_F(\rho)\in[0,1]$ but also it has a geometric interpretation.
 
\noindent Another intuitive measure of coherence is the $l_1$-norm measure of coherence defined as \cite{TBaumgratz}
 \begin{equation}\label{1.3}
 C_{l_1}(\rho)=\sum_{\substack{i,j=1,\\i\neq j}}^d|\rho_{ij}|
 \end{equation}
where sum over all of the off-diagonal terms are considered. One can recast the above matrix in a form as
\begin{equation}\label{1.4}
C_{l_1}(\rho)=\sum_{i,j=1}^d|\rho_{ij}-\delta_{ij}|.
\end{equation}
We can see that the definitions in eq.(\ref{1.3}) and eq.(\ref{1.4}) are equivalent definitions for the coherence measure. The definition in eq.(\ref{1.4}) is more intuitive in a sense that it measures the distance of the elements of the density matrix from the elements of the $d$ dimensional identity matrix which describes the matrix for completely incoherent state (as also has been discussed for the Frobenius-norm measure of coherence case).

\noindent Another very robust measure of coherence is the relative entropy of coherence defined as \cite{TBaumgratz}
\begin{equation}\label{1.5}
C_{\text{Rel}}(\rho)=S(\rho_{\text{diag}})-S(\rho)
\end{equation}
where the maximum possible value of coherence in a state is given by $\ln d$ with $S(\rho)=-\text{tr}[\rho\ln\rho]$ giving the von-Neumann entropy for the density matrix $\rho$ and $\rho_\text{diag}$ denoting the density matrix constructed using the diagonal elements of the density matrix $\rho$. Finally, involving the off-diagonal elements of a density matrix, one can construct another measure of coherence, also known as the $l_2$ norm of coherence, as \cite{TBaumgratz}
\begin{equation}
C_{l_2}(\rho)\equiv\sum_{\substack{i,j=1,\\i\neq j}}^d|\rho_{ij}|^2~.
\end{equation}
As has been argued in \cite{TBaumgratz}, the above measure of coherence does not construct a valid coherence monotone. Although for the sake of completeness, we have still calculated the measure of coherence for the different cases considered in this analysis.

\noindent   The construction of our paper goes as follows. In section (\ref{2}), we have reviewed the basic representation of the single-particle quantum state under Lorentz boost. In section (\ref{3}), we analyze the coherence of a spin-$\frac{1}{2}$ particle in $(1+1)$ dimensions and calculated different measures of coherence. In section (\ref{4}) we have considered the coherence of a spin-$\frac{1}{2}$ particle in $(3+1)$-dimensions. . In contrast to the electron, the neutron has a narrow-uncertainty wave packet. In section (\ref{5}), we provide some plots considering the results in earlier sections. We summarize our results in section (\ref{6}). 
\section{Representation of the single-particle quantum state under Lorentz boost}
\label{2}
\noindent In the domain of relativistic quantum mechanics one of the most pertinent questions is how a quantum state behaves under Lorentz boosts. To see the effects of Lorentz boosts on the quantum states at first, we consider a single particle quantum state $\vert \Psi \rangle$. It is well known that an inertial reference frame $\mathcal{O}$ is related to another inertial frame $\mathcal{O'}$ through a Poincare transformation
\begin{eqnarray}
x'^{\mu}=T(\Lambda,a)x^{\nu}=\Lambda^{\mu}_{\hspace{0.2cm}\nu}\,x^{\nu}+a^{\mu}
\end{eqnarray}
where $x=(x^{0},\,\textbf{x})$ is the spacetime coordinates of $\mathcal{O}$, and similarly for $\mathcal{O'}$. $T(\Lambda,a)$ induces a unitary transformation on quantum states as
\begin{eqnarray}
\vert \Psi \rangle \rightarrow \mathcal{U}(\Lambda,a)\vert \Psi \rangle\,.
\end{eqnarray}
Now we can define the single particle quantum state as 
\begin{eqnarray}
\vert \Psi \rangle =\vert \mathbf{p}\rangle \otimes \vert \sigma \rangle \equiv \vert \mathbf{p},\sigma\rangle
\end{eqnarray}
where $\mathbf{p}$ labels the spatial components of the  4-momentum $p^{\mu}$ with $p^0=\sqrt{\mathbf{p}^2+m^2}$ and $\sigma$ labels the spin for massive particles. The quantum states $\vert p,\sigma\rangle$ are eigenvectors of the momentum operator $P^{\mu}$ with eigenvalues $p^{\mu}$ and satisfies the relation
\begin{eqnarray}
P^{\mu}\vert \mathbf{p},\sigma\rangle =p^{\mu}\vert \mathbf{p},\sigma\rangle\,.
\end{eqnarray}

\noindent Under pure Lorentz boosts $\Lambda$, a single particle quantum state $\vert \mathbf{p},\sigma \rangle$ transforms as\footnote{A detailed derivation of eq.(\ref{t-state}) has been presented in an Appendix.}
\begin{eqnarray}
\mathcal{U}(\Lambda)\vert \mathbf{p},\sigma \rangle=\sqrt{\frac{(\Lambda p)^0}{p^0}}\displaystyle\sum_{\sigma'}\mathcal{D}_{\sigma'\sigma}(W(\Lambda,\mathbf{p}))\vert \Lambda \mathbf{p},\sigma' \rangle\,.\label{t-state}\nonumber\\
\end{eqnarray}
In the above equation $\Lambda\mathbf{p}$ is the spatial part of the momentum after Lorentz transformation and $W(\Lambda,\mathbf{p})$ denotes one of the elements of the unitary representation of little group $\mathcal{D}(W(\Lambda,\mathbf{p}))$.
For massive particles, the little group is the rotation group in three dimensions and hence $W(\Lambda,\mathbf{p})\in SO(3)$. However, its unitary representation takes the form $\mathcal{D}(W(\Lambda,\mathbf{p}))\in SU(2)$.

\noindent We consider the four-momentum of the particle in the form
\begin{eqnarray}
p^{\mu}=(m \cosh \zeta, m \sinh \zeta \hat{f})\,
\end{eqnarray}
where $m$ is the mass of the particle, and the velocity of the moving frame $\mathcal{O'}$, $\vec{v}=\tanh\alpha \,\hat{e}$. Then the unitary representation of the little group $\mathcal{D}(W(\Lambda,\mathbf{p}))$ takes the form \cite{RiddhiASM}
\begin{eqnarray}
\mathcal{D}(W(\Lambda,\mathbf{p}))=\cos \frac{\phi}{2}\mathds{1}+i\sin\frac{\phi}{2}(\Sigma\cdot\hat{n})
\end{eqnarray}
where
\begin{eqnarray}
\cos\frac{\phi}{2}&=&\frac{\cosh\frac{\alpha}{2}\cosh\frac{\zeta}{2}+\sinh\frac{\alpha}{2}\sinh\frac{\zeta}{2}(\hat{e}\cdot\hat{f})}{\sqrt{\frac{1}{2}+\frac{1}{2}\cosh\alpha\cosh\zeta+\frac{1}{2}\sinh\alpha\sinh\zeta(\hat{e}\cdot\hat{f})}}\nonumber\\
&\,&\\
\sin\frac{\phi}{2}\hat{n}&=&\frac{\sinh\frac{\alpha}{2}\sinh\frac{\zeta}{2}(\hat{e}\times\hat{f})}{\sqrt{\frac{1}{2}+\frac{1}{2}\cosh\alpha\cosh\zeta+\frac{1}{2}\sinh\alpha\sinh\zeta(\hat{e}\cdot\hat{f})}}\nonumber\\
&\,&
\end{eqnarray}
with $\phi\text{ and }\hat{n}$ being the angle and axis of the Wigner rotation.

\noindent In an inertial frame $\mathcal{O}$, a pure state can be written as
\begin{eqnarray}
\vert \Psi\rangle &=&\displaystyle\sum_{\sigma}\int d^3\mathbf{p}\, \psi(\mathbf{p})\vert \mathbf{p}\rangle\otimes c_{\sigma}\vert \sigma\rangle\nonumber\\
&=&\displaystyle\sum_{\sigma}\int d^3\mathbf{p}\, \psi(\mathbf{p})c_{\sigma}\vert \mathbf{p},\sigma\rangle\,.\label{psi}
\end{eqnarray}
The density matrix of the corresponding state becomes
\begin{eqnarray}
\varrho&=&\vert\Psi\rangle\langle\Psi\vert\nonumber\\
&=&\displaystyle\sum_{\sigma_1,\sigma_2}\int d^3\mathbf{p}_{1}\int d^3\mathbf{p}_{2}\, \psi(\mathbf{p}_{1})\psi^{\ast}(\mathbf{p}_{2})c_{\sigma_1}c^{\ast}_{\sigma_2}\nonumber\\&\times&\vert \mathbf{p}_{1},\sigma_{1}\rangle\langle \mathbf{p}_{2},\sigma_{2}\vert\nonumber~.\\\label{varrho}
\end{eqnarray}
Taking partial trace over the momentum degrees of freedom, the reduced density matrix for the spin reads
\begin{eqnarray}
\rho&=&\text{Tr}_p[\vert\Psi\rangle\langle\Psi\vert]=\int d^3\mathbf{p}\langle \mathbf{p}\vert\Psi\rangle\langle\Psi\vert\mathbf{p}\rangle\nonumber\\
&=&\displaystyle\sum_{\sigma_1,\sigma_2}\int d^3\mathbf{p}\int d^3\mathbf{p}_{1}\int d^3\mathbf{p}_{2}\,\delta^{(3)}(\mathbf{p}-\mathbf{p}_{1})\,\delta^{(3)}(\mathbf{p}-\mathbf{p}_{2}) \nonumber\\
&\,\, &\,\times\,\psi(\mathbf{p}_{1})\psi^{\ast}(\mathbf{p}_{2})c_{\sigma_1}c^{\ast}_{\sigma_2}\vert\sigma_{1}\rangle\langle\sigma_{2}\vert\nonumber\\
&=&\displaystyle\sum_{\sigma_1,\sigma_2}\int d^3\mathbf{p}\,\psi(\mathbf{p})\psi^{\ast}(\mathbf{p})c_{\sigma_1}c^{\ast}_{\sigma_2}\vert\sigma_{1}\rangle\langle\sigma_{2}\vert\,.\label{rho}
\end{eqnarray} 
With respect to a moving frame $\mathcal{O'}$ which is connected with the inertial frame $\mathcal{O}$ through a Lorentz transformation $\Lambda$, this state can be represented as
\begin{eqnarray}
\vert \Psi_{\Lambda}\rangle&=&\displaystyle\sum_{\sigma}\int d^3\mathbf{p}\, \psi(\mathbf{p})c_{\sigma}\sqrt{\frac{(\Lambda p)^0}{p^0}}\displaystyle\nonumber\\&\times&\sum_{\sigma'}\mathcal{D}_{\sigma'\sigma}(W(\Lambda,\mathbf{p}))\vert \Lambda \mathbf{p},\sigma' \rangle\,.\nonumber\\\label{boostedpsi}
\end{eqnarray}
Therefore, the density matrix in this boosted reference frame is given by
\begin{eqnarray}
\varrho_{\Lambda}&=&\displaystyle\sum_{\sigma_1,\sigma_2}\int d^3\mathbf{p}_{1}\int d^3\mathbf{p}_{2}\,\sqrt{\frac{(\Lambda p_1)^0(\Lambda p_2)^0}{p_1^0p_2^0}} \psi(\mathbf{p}_{1})\psi^{\ast}(\mathbf{p}_{2})\nonumber\\
&\times & c_{\sigma_1}c^{\ast}_{\sigma_2}\displaystyle\sum_{\sigma'_1,\sigma'_2}\mathcal{D}_{\sigma'_1\sigma_1}(W(\Lambda,\mathbf{p}_1))\vert \Lambda\mathbf{p}_{1},\sigma'_{1}\rangle\langle \Lambda\mathbf{p}_{2},\sigma'_{2}\vert\nonumber\\
&\times&\mathcal{D}_{\sigma'_2\sigma_2}^{\dagger} (W(\Lambda,\mathbf{p}_2))~.\label{boostedvarrho}
\end{eqnarray} 
Hence, the corresponding reduced density matrix is given by
\begin{eqnarray}
\rho_{\Lambda}&=&\text{Tr}_p[\vert\Psi_{\Lambda}\rangle\langle\Psi_{\Lambda}\vert]\nonumber\\
&=&\displaystyle\sum_{\sigma_1,\sigma_2,\sigma'_1,\sigma'_2}\int d^3\mathbf{p}\int d^3\mathbf{p}_{1}\int d^3\mathbf{p}_{2}\,\sqrt{\frac{(\Lambda p_1)^0(\Lambda p_2)^0}{p_1^0p_2^0}}\nonumber\\
&\times & \delta^{(3)}(\mathbf{p}-\Lambda\mathbf{p}_{1})\,\delta^{(3)}(\mathbf{p}-\Lambda\mathbf{p}_{2})\mathcal{D}_{\sigma'_1\sigma_1}(W(\Lambda,\mathbf{p}_1)) \nonumber\\
&\times & \psi(\mathbf{p}_{1})\psi^{\ast}(\mathbf{p}_{2})c_{\sigma_1}c^{\ast}_{\sigma_2}\vert\sigma'_{1}\rangle\langle\sigma'_{2}\vert\mathcal{D}_{\sigma'_2\sigma_2}^{\dagger}(W(\Lambda,\mathbf{p}_2))\nonumber\\
&=&\displaystyle\sum_{\sigma_1,\sigma_2,\sigma'_1,\sigma'_2}\int d^3\mathbf{p}\,\psi(\mathbf{p})\psi^{\ast}(\mathbf{p})c_{\sigma_1}c^{\ast}_{\sigma_2}\mathcal{D}_{\sigma'_1\sigma_1}(W(\Lambda,\mathbf{p}))\nonumber\\
&\times &\vert\sigma'_{1}\rangle\langle\sigma'_{2}\vert\mathcal{D}_{\sigma'_2\sigma_2}^{\dagger}(W(\Lambda,\mathbf{p}))\,,\label{boostedrho}
\end{eqnarray}
where we have used
\begin{eqnarray}
\delta^{(3)}(\Lambda\mathbf{p}_{1}-\Lambda\mathbf{p}_{2})=\frac{p_1^0}{(\Lambda p_1)^0}\delta^{(3)}(\mathbf{p}_{1}-\mathbf{p}_{2})\,.
\end{eqnarray}
Here we observe that the state in eq.\eqref{psi} and the density matrix in eq.\eqref{varrho} is separable in spin and momentum, however, in eq.(s)(\ref{boostedpsi}, \ref{boostedvarrho}) spin and momentum has been coupled due to momentum dependent rotation and is known as spin-momentum entanglement \cite{RiddhiASM,Weinberg}.
\section{DIfferent measure of coherence for spin 1/2 particles under Lorentz boost in (1+1)-dimensions}
\label{3}
\noindent In this section, we shall obtain the Frobenius-norm measure of coherence for spin 1/2 particles with spin-momentum entanglement.
 The state of a single spin-1/2 particle going in the $\hat{x}$-direction with respect to an inertial observer $\mathcal{O}$ is given by
\begin{equation}\label{psi1/2}
|\Psi\rangle=\frac{1}{\sqrt{2}}\int dp_x\psi(p_x)|p_x\rangle\otimes\left(|0\rangle+|1\rangle\right)
\end{equation}
with $p^{\mu}=(m \cosh \zeta, m \sinh \zeta \hat{x})$ being the two-momentum of the particle and $m$ being the mass of the particle. The density matrix corresponding to this state $\vert\Psi\rangle$ is therefore given by
\begin{equation}\label{varrho(1+1)}
\rho=\frac{1}{2}\int dp_{x_1}\int dp_{x_2}\psi(p_{x_1})\psi^*(p_{x_2})|p_{x_1}\rangle\langle p_{x_2}|\otimes(\mathbb{1}+\Sigma_1)~.
\end{equation}
Hence, the reduced density matrix is given by
\begin{equation}\label{rho(1+1)}
\rho=\frac{1}{2}\int dp_x\psi(p_x)\psi^*(p_x)(\mathbb{1}+\Sigma_1)~.
\end{equation}
           In the frame of $\mathcal{O'}$ moving with velocity $\vec{v}=\tanh\alpha\,\hat{z}$, the state of the particle is given by
\begin{equation}
\begin{split}
\vert \Psi_{\Lambda}\rangle&=\frac{1}{\sqrt{2}}\int dp_x \psi(p_x)\sqrt{\frac{(\Lambda p)^0}{p^0}}\mathcal{D}(W(\Lambda,p_x))(\vert \Lambda p_x,0 \rangle\\&+\vert \Lambda p_x,1\rangle)\,,
\end{split}
\label{boostedpsi1/2}
\end{equation}
where
\begin{equation}
\mathcal{D}(W(\Lambda,p_x))=\cos \frac{\phi}{2}\mathds{1}+i\sin\frac{\phi}{2}(\Sigma_2)\,\label{D(1+1)}
\end{equation}
with
\begin{eqnarray}
\cos\frac{\phi}{2}&=&\frac{\cosh\frac{\alpha}{2}\cosh\frac{\zeta}{2}}{\sqrt{\frac{1}{2}+\frac{1}{2}\cosh\alpha\cosh\zeta}}\nonumber\\
\sin\frac{\phi}{2}&=&\frac{\sinh\frac{\alpha}{2}\sinh\frac{\zeta}{2}}{\sqrt{\frac{1}{2}+\frac{1}{2}\cosh\alpha\cosh\zeta}}\label{comp(1+1)}~.
\end{eqnarray}
Here we have employed $\hat{z}\cdot\hat{x}=0\text{ and } \hat{z}\times\hat{x}=\hat{y}$. As $\hat{n}=\hat{y}$ with $\hat{n}$ denoting the alignment of the direction of the axis about which the Wigner rotation takes place. It is quite straightforward to say that the axis of rotation is directed along the $y$ axis.\\
Substituting eq.(s)(\ref{D(1+1)}, \ref{comp(1+1)}) in eq.\eqref{boostedpsi1/2}, we get
\begin{eqnarray}
\vert \Psi_{\Lambda}\rangle=&\frac{1}{\sqrt{2}}&\int dp_x\, \sqrt{\frac{(\Lambda p)^0}{p^0}}\psi(p_x)\Bigg[\left(\cos\frac{\phi}{2}+\sin\frac{\phi}{2}\right)\Bigg.\nonumber\\
&\times&\Bigg.\vert \Lambda p_x,0 \rangle+\left(\cos\frac{\phi}{2}-\sin\frac{\phi}{2}\right)\vert \Lambda p_x,1\rangle\Bigg]\,.\label{boostedpsi1/2-1}
\end{eqnarray}
The density matrix corresponding to the above state is given by
\begin{equation}
\begin{split}
\varrho_{\Lambda}&=\frac{1}{2}\int dp_{x_{1}}\int dp_{x_2}\,\sqrt{\frac{(\Lambda p_1)^0(\Lambda p_2)^0}{p_1^0p_2^0}} \psi(p_{x_1})\psi^{\ast}({p}_{x_2})\\
&\times \bigg[\mathcal{A}_{1}\mathcal{A}_{2}\vert \Lambda {p}_{x_1},0\rangle\langle \Lambda {p}_{x_2},0\vert+\mathcal{A}_{1}\mathcal{B}_{2}\vert \Lambda{p}_{x_1},0\rangle\langle \Lambda{p}_{x_2},1\vert\bigg.\\
&\times \bigg. \mathcal{A}_{2}\mathcal{B}_{1}\vert \Lambda{p}_{x_1},1\rangle\langle \Lambda{p}_{x_2},0\vert+\mathcal{B}_{1}\mathcal{B}_{2}\vert \Lambda{p}_{x_1},1\rangle\langle \Lambda{p}_{x_2},1\vert\bigg]\label{varrhospin1/2}
\end{split}
\end{equation}
where $\mathcal{A}_{i}=\left(\cos\frac{\phi_{i}}{2}+\sin\frac{\phi_{i}}{2}\right)$ and $\mathcal{B}_{i}=\left(\cos\frac{\phi_{i}}{2}-\sin\frac{\phi_{i}}{2}\right);i=1,2$. 

\noindent The reduced density matrix corresponding to $\varrho_{\Lambda}$ is given by
\begin{eqnarray}
\rho_{\Lambda}
=&\frac{1}{2}&\int d p_x\,\psi(p_x)\psi^{\ast}(p_x)\big[\mathcal{A}^{2}\vert 0\rangle\langle 0\vert+\mathcal{A}\mathcal{B}\bigl(\vert 0\rangle\langle 1\vert\big.\bigl.\nonumber\\
&\,& \big. \bigr.+\vert 1\rangle\langle 0\vert\bigr)+\mathcal{B}^{2}\vert 1\rangle\langle 1\vert\big]\label{rhospin1/2}
\end{eqnarray}
One can now consider particular forms of $\psi(p_x)$ to calculate the reduced density $\rho_{\Lambda}$.\\
Let us consider $\psi(p_x)$ in the following form\footnote{If one starts with a three dimensional construction where $\rho_{\Lambda}
=\frac{1}{2}\int d^3\mathbf{p}\,\psi(\mathbf{p})\psi^{\ast}(\mathbf{p})\big[\mathcal{A}^{2}\vert 0\rangle\langle 0\vert+\mathcal{A}\mathcal{B}\bigl(\vert 0\rangle\langle 1\vert\big.\bigl.
\, \big. \bigr.+\vert 1\rangle\langle 0\vert\bigr)+\mathcal{B}^{2}\vert 1\rangle\langle 1\vert\big]$, then one needs to just consider $\psi(\mathbf{p})=\psi(p_x)\delta(p_y)\delta(p_z)=f(p_x)\delta(p_y)\delta(p_z)$. This particular choice of  $\psi(\mathbf{p})$ restores the form of the reduced density matrix in eq.(\ref{rhospin1/2}).}
\begin{eqnarray}
\psi(p_x)=f(p_x)\label{psipx}
\end{eqnarray}
with $f(p_x)$ being given by
\begin{eqnarray}
f(p_x)=\frac{1}{\sqrt{\sigma^{2n+1}\Gamma(n+\frac{1}{2})}}p_x^{n}\,\mathrm{e}^{-\frac{1}{2}\left(\frac{p_x}{\sigma}\right)^2}\label{fx}
\end{eqnarray}
where $n\in\mathbb{R}$.
Note that the case corresponding to $n=0$ was investigated in \cite{RiddhiASM}. In this work, we consider a more general form of the function $f(p_x)$.
Using eq.(s)(\ref{psipx}, \ref{fx}) in eq.\eqref{rhospin1/2}, we get
\begin{eqnarray}
\rho_{\Lambda}
=&\frac{1}{2}&\int_{-\infty}^{+\infty} dp_x\,\vert f(p_x)\vert^2\big[\mathcal{A}^{2}\vert 0\rangle\langle 0\vert+\mathcal{A}\mathcal{B}\bigl(\vert 0\rangle\langle 1\vert\big.\bigl.\nonumber\\
&\,& \big. \bigr.+\vert 1\rangle\langle 0\vert\bigr)+\mathcal{B}^{2}\vert 1\rangle\langle 1\vert\big]~.
\end{eqnarray}
Taking $\sinh\zeta=\frac{p_x}{m}, \cosh\zeta=\sqrt{1+\left(\frac{p_x}{m}\right)^2},\sinh\alpha=a,\text{ and }\cosh\alpha=b$, we get
\begin{eqnarray}
\mathcal{A}^2&=&1+\frac{a\,\frac{p_x}{m}}{1+b\,\sqrt{1+\left(\frac{p_x}{m}\right)^2}}\nonumber\\
\mathcal{B}^2&=&1-\frac{a\,\frac{p_x}{m}}{1+b\,\sqrt{1+\left(\frac{p_x}{m}\right)^2}}\\
\mathcal{A\,B}&=&\frac{b+\sqrt{1+\left(\frac{p_x}{m}\right)^2}}{1+b\,\sqrt{1+\left(\frac{p_x}{m}\right)^2}}~.\nonumber
\end{eqnarray}
Therefore, the components of the reduced density matrix $\rho_{\Lambda}$ reads
\begin{eqnarray}
\rho_{\Lambda_{00}}&=&\frac{1}{2}\int_{-\infty}^{+\infty} dp_x\,\vert f(p_x)\vert^2\left(1+\frac{a\,\frac{p_x}{m}}{1+b\,\sqrt{1+\left(\frac{p_x}{m}\right)^2}}\right)\,,\nonumber\\
\rho_{\Lambda_{11}}&=&\frac{1}{2}\int_{-\infty}^{+\infty} dp_x\,\vert f(p_x)\vert^2\left(1-\frac{a\,\frac{p_x}{m}}{1+b\,\sqrt{1+\left(\frac{p_x}{m}\right)^2}}\right)\,,\\
\rho_{\Lambda_{01}}&=&\rho_{\Lambda_{10}}=\frac{1}{2}\int_{-\infty}^{+\infty} dp_x\,\vert f(p_x)\vert^2\left(\frac{b+\sqrt{1+\left(\frac{p_x}{m}\right)^2}}{1+b\,\sqrt{1+\left(\frac{p_x}{m}\right)^2}}\right)\,.\nonumber
\end{eqnarray}
We first start by considering the $\rho_{\Lambda_{00}}$ element.
\begin{equation}\label{rho00}
\begin{split}
\rho_{\Lambda_{00}}&=\frac{1}{2}\int_{-\infty}^{+\infty}dp_x|f(p_x)|^2\left(1+\frac{a\,\frac{p_x}{m}}{1+b\,\sqrt{1+\left(\frac{p_x}{m}\right)^2}}\right)\\
&=\frac{1}{2}\int_{-\infty}^{+\infty}dp_x|f(p_x)|^2\\
&=\frac{1}{2\Gamma\left(n+\frac{1}{2}\right)}\int_{-\infty}^{+\infty}d\left(\frac{p_x}{\sigma}\right)\left(\frac{p_x}{\sigma}\right)^{2n}e^{-\left(\frac{p_x}{\sigma}\right)^2}\\
&=\frac{1}{2\Gamma\left(n+\frac{1}{2}\right)}\int_0^\infty~d\zeta_x \zeta_x^{n-\frac{1}{2}}~e^{-\rho_x}
\end{split}
\end{equation}
where $\zeta_x=\frac{p_x^2}{\sigma^2}$. The above integral is convergent provided $n>-\frac{1}{2}$. Hence, for $n>-\frac{1}{2}$, we can obtain the form of $\rho_{\Lambda_{00}}$ element as
\begin{equation}\label{rho00.1}
\rho_{\Lambda_{00}}=\frac{1}{2\Gamma\left(n+\frac{1}{2}\right)}\Gamma\left(n+\frac{1}{2}\right)=\frac{1}{2}~.
\end{equation}
Similarly, one can obtain the $\{1,1\}$ element of the reduced density matrix as $\rho_{\Lambda_{11}}=\frac{1}{2}$. It is now no more possible to obtain the analytical form of the off-diagonal elements of the reduced density matrix. Hence, we consider the width of the wave function to be small enough compared to the mass of the particle, that is $\left(\frac{\sigma}{m}\right)\ll1$. This now helps us to obtain the off-diagonal components of the density matrix up to $\mathcal{O}\left(\left(\frac{\sigma}{m}\right)^2\right)$ as
\begin{eqnarray}
\rho_{\Lambda_{00}}&=&\rho_{\Lambda_{11}}=\frac{1}{2}\\
\rho_{\Lambda_{01}}&=&\rho_{\Lambda_{10}}=\frac{1}{2}-\frac{2n+1}{8}\left(\frac{\cosh\alpha-1}{\cosh\alpha+1}\right)\left(\frac{\sigma}{m}\right)^2~.\nonumber\label{rhocomp(1+1)}
\end{eqnarray}
For $n=0$, the above results reduce to those given in \cite{RiddhiASM}. Our results show the effect of $n$ coming through the generalized wave packet that we have considered in this work. The matrix form of the reduced density matrix can therefore be written as
\begin{eqnarray}
\rho_{\Lambda}=\begin{pmatrix}
\frac{1}{2}&\frac{1}{2}-\mathcal{F}\\\frac{1}{2}-\mathcal{F}&\frac{1}{2}\label{rhomat}
\end{pmatrix}
\end{eqnarray}
with $\mathcal{F}=\frac{2n+1}{8}\left(\frac{\cosh\alpha-1}{\cosh\alpha+1}\right)\left(\frac{\sigma}{m}\right)^2$. \\
\subsection{Frobenius-norm measure of coherence}\label{3.a}
\noindent With the above result in hand, we would now like to obtain the Frobenius-norm measure of coherence for the above reduced density matrix. The Frobenius-norm measure of coherence is defined as \cite{BID}
\begin{eqnarray}
C_{F}(\rho)\equiv\sqrt{\frac{d}{d-1}\displaystyle\sum_{i=1}^d \left(\lambda_{i}-\frac{1}{d}\right)^2}\label{co_me}
\end{eqnarray}
where $d$ is the dimension of the Hilbert space and $\{\lambda_i\}$ are the eigenvalues of the density matrix $\rho$. Hence we need to first calculate the eigenvalues of $\rho_\Lambda$ (in eq.(\ref{rhomat})). Calculating the eigenvalues of the reduced density matrix $\rho_{\Lambda}$ from the eq.\eqref{rhomat} and using it in eq.\eqref{co_me}, we get
\begin{eqnarray}
C_{F}(\rho_{\Lambda})=\left[1-\frac{2n+1}{4}\left(\frac{\cosh\alpha-1}{\cosh\alpha+1}\right)\left(\frac{\sigma}{m}\right)^2\right]\label{co_me(1+1)}~.
\end{eqnarray}
Now as $\alpha$ is the rapidity parameter for the boosted observer, therefore we can write
\begin{eqnarray}
\cosh\alpha=\gamma\equiv\frac{1}{\sqrt{1-(\frac{v}{c})^2}}
\end{eqnarray}
where $v$ is the velocity of the boosted observer with respect to the inertial one. Therefore, when $v\rightarrow 0$, $\cosh\alpha\rightarrow 1$. In this limit, eq.\eqref{co_me(1+1)} becomes
\begin{eqnarray}
C_{F}(\rho_{\Lambda})=1\label{co_me(1+1)1}\,.
\end{eqnarray}
On the other hand, when $v\rightarrow c,~\text{ then }\cosh\alpha\rightarrow \infty$. In this limit, eq.\eqref{co_me(1+1)} becomes
\begin{eqnarray}
C_{F}(\rho_{\Lambda})=\left[1-\frac{2n+1}{4}\left(\frac{\sigma}{m}\right)^2\right]\label{co_me(1+1)2}\,.
\end{eqnarray}
Since $C_{F}(\rho_{\Lambda})\geq0$, hence we get an upper bound of the parameter $n$ to be
\begin{eqnarray}
n\leq \left[2\left(\frac{m}{\sigma}\right)^2-\frac{1}{2}\right]~.\label{bound_on_n}
\end{eqnarray}
Eq.(s)(\ref{co_me(1+1)2},\ref{bound_on_n}) are some of the main findings in this work. We can further obtain a lower bound to $n$ by claiming that the measure of coherence at all times must be less than or equal to unity. As we are considering $\frac{\sigma}{m}<1$, we cannot use the equality condition. Hence, using eq.(\ref{co_me(1+1)2}), we obtain
\begin{equation}\label{co_melb}
\begin{split}
&1-\frac{2n+1}{4}\left(\frac{\sigma}{m}\right)^2<1\\
\implies&n> -\frac{1}{2}~.
\end{split}
\end{equation}
It is important to note that this lower bound is same as the condition required to perform the integration in eq.(\ref{rho00}). This is an excellent agreement between the mathematical and physical requirements of the system in consideration.
Combining eq.(\ref{bound_on_n}) with eq.(\ref{co_melb}), we obtain a range for the parameter $n$ as
\begin{equation}\label{n_range}
-\frac{1}{2}< n\leq\left[2\left(\frac{m}{\sigma}\right)^2-\frac{1}{2}\right]~.
\end{equation}
One can also obtain the minimum upperbound for $n$ when $\sigma\simeq m$ as $n\simeq 1.5$. This shows that the maximum value of $n$ that one can consider for a generalized Gaussian wave-packet completely depends on the chosen width of the Gaussian wave-packet. Eq.(s)(\ref{co_me(1+1)2},\ref{n_range}) are very important findings for the (1+1)-dimensional analysis.
\subsection{$l_1$-norm measure of coherence}\label{3.b}
\noindent In this subsection, we shall calculate the $l_1$ norm measure of coherence which is defined as \cite{TBaumgratz}
\begin{equation}\label{l1.1}
C_{l_1}=\sum\limits_{\substack{i,j=0\\i\neq j}}^1\left|\rho_{ij}\right|=\left|\rho_{01}\right|+\left|\rho_{10}\right|.
\end{equation}
From the forms of the elements of the density matrix in eq.(\ref{rhocomp(1+1)}), it is easy to check that the off-diagoanl components of the density matrix are equal, $\rho_{\Lambda_{01}}=\rho_{\Lambda_{10}}$. Hence, the $l_1$-norm measure of coherence reads
\begin{equation}\label{l1.2}
\begin{split}
C_{l_1}(\rho_\Lambda)&=2|\rho_{\Lambda_{01}}|\\&=2\left[\frac{1}{2}-\frac{2n+1}{8}\left[\frac{\cosh\alpha-1}{\cosh\alpha+1}\right]\left(\frac{\sigma}{m}\right)^2\right]\\
&=1-\frac{2n+1}{4}\left[\frac{\cosh\alpha-1}{\cosh\alpha+1}\right]\left(\frac{\sigma}{m}\right)^2.
\end{split}
\end{equation}
The above result is very interesting in the sense that for this case the Frobenius norm measure is exactly identical to the $l_1$ norm measure of coherence. It should be noted that the $l_1$ norm of coherence is considered as the best measure for quantifying coherence. We next move towards calculating the $l_2$ norm of coherence. In order to investigate the reason behind the two measure of coherence to be equal, we start by defining a density matrix as
\begin{equation}\label{l1.3}
\rho_\Lambda=
\begin{pmatrix}
\frac{1}{2}&&\rho_{\Lambda_{01}}\\
\rho_{\Lambda_{10}}&&\frac{1}{2}
\end{pmatrix}~.
\end{equation}
The eigenvalues are obtained as
\begin{equation}\label{l1.4}
\lambda_{\pm}=\frac{1}{2}\pm\sqrt{\rho_{\Lambda_{01}}\rho_{\Lambda_{10}}}~.
\end{equation}
The Frobenius norm measure of coherence for $d=2$ from eq.(\ref{co_me}) reads
\begin{equation}\label{l1.5}
\begin{split}
C_F(\rho)&=\sqrt{2\left(\left(\lambda_+-\frac{1}{2}\right)^2+\left(\lambda_--\frac{1}{2}\right)^2\right)}\\
&=\sqrt{2(\rho_{\Lambda_{01}}\rho_{\Lambda_{10}}+\rho_{\Lambda_{01}}\rho_{\Lambda_{10}})}\\
\implies C_F(\rho)&=2\sqrt{\rho_{\Lambda_{01}}\rho_{\Lambda_{10}}}~.
\end{split}
\end{equation}
The $l_1$ norm of coherence on the other hand just reads
\begin{equation}\label{l1.6}
\begin{split}
C_{l_1}(\rho)=|\rho_{\Lambda_{01}}|+|\rho_{\Lambda_{10}}|.
\end{split}
\end{equation}
Eq.(\ref{l1.6}) is equal to eq.(\ref{l1.5}) provided that
\begin{equation}\label{l1.7}
\rho_{\Lambda_{01}}=\rho^*_{{\Lambda}_{10}}
\end{equation} 
and the diagonal elements are $\frac{1}{2}$. In the present case we can observe from eq.(\ref{rhomat}) that $\rho_{\Lambda_{01}}=\rho_{\Lambda_{10}}$ with the diagonal elements being $\frac{1}{2}$. 
The fact that the Frobenius norm and $l_1$ norm matches exactly is due to the reason that the structure of the reduced density matrix (eq.(\ref{rhomat})) is exactly similar to the form of the density matrix defined in eq.(\ref{l1.3}) where the off diagonal real components are equal to each other.
\subsection{$l_2$ norm of coherence}\label{3.c}
\noindent Another important measure of coherence is the $l_2$-norm of coherence. Although it gives convincing outcomes in several scenarios, it is often argued to be not a very good measure of coherence \cite{TBaumgratz}. We have calculated this coherence measure just for the sake of completeness of our analysis. The $l_2$ norm of coherence is defined as \cite{TBaumgratz}
\begin{equation}\label{l2.1}
\begin{split}
C_{l_2}(\rho_\Lambda)\equiv\sum\limits_{\substack{i,j=0\\i\neq j}}^2|\rho_{ij}|^2=|\rho_{01}|^2+|\rho_{10}|^2.
\end{split}
\end{equation}
Using the form of the density matrix in eq.(\ref{rhomat}), the $l_2$ norm measure of coherence reads
\begin{equation}\label{l2.2}
\begin{split}
C_{l_2}(\rho_\Lambda)&=\left(\frac{1}{2}-\mathcal{F}\right)^2+\left(\frac{1}{2}-\mathcal{F}\right)^2\simeq\frac{1}{2}-2\mathcal{F}\\
&=\frac{1}{2}-\frac{2n+1}{4}\left[\frac{\cosh\alpha-1}{\cosh\alpha+1}\right]\left(\frac{\sigma}{m}\right)^2~.
\end{split}
\end{equation}
\subsection{Relative entropy of coherence}\label{3.d}
\noindent Another, quite simple analytical measure of coherence is often given by the relative entropy of coherence which is defined as \cite{TBaumgratz}
\begin{equation}\label{d.1}
C_{\text{Rel}}(\rho_\Lambda)=S(\rho_{\text{diag}})-S(\rho)
\end{equation}
where $\rho_{\text{diag}}=\sum_{i}\rho_{ii}|i\rangle\langle i|$ with $S(\rho_{\text{diag}})$ being defined as
\begin{equation}\label{d.2}
\begin{split}
S(\rho_{\text{diag}})&=-\text{tr}\left[\rho_{\text{diag}}\ln\left[\rho_{\text{diag}}\right]\right]\\
&=-\sum_{i=1}^2 \lambda^{\text{diag}}_i\ln\lambda^{\text{diag}}_i\\
&=-(\lambda^{\text{diag}}_1\ln\lambda^{\text{diag}}_1+\lambda^{\text{diag}}_2\ln\lambda^{\text{diag}}_2)
\end{split}
\end{equation}
where $\lambda^{\text{diag}}_i$ (for $i\in\{1,2\}$) denotes the eigenvalues of the diagonal reduced density matrix. The von-Neumann entropy corresponding to the reduced density matrix $\rho$ takes the form
\begin{equation}\label{d.3}
\begin{split}
S(\rho)&=-\text{tr}[\rho\ln\rho]\\
&=-\sum_{i=1}^{2}\lambda_i\ln\lambda_i\\
&=-\left(\lambda_1\ln\lambda_1+\lambda_2\ln\lambda_2\right)
\end{split}
\end{equation}
where $\lambda_i$ (for $i\in\{1,2\}$) denotes the eigenvalues of the reduced density matrix. For the reduced density matrix in eq.(\ref{rhomat}), we obtain the eigenvalues corresponding to the diagonal reduced density matrix ($\rho_\Lambda^{\text{diag}}$) as
\begin{equation}\label{d.4}
\lambda_{1}^{\text{diag}}=\lambda_2^\text{diag}=\frac{1}{2}~.
\end{equation}
The eigenvalues corresponding to the reduced density matrix in eq.(\ref{rhomat}) as
\begin{equation}\label{d.5}
\lambda_1=\mathcal{F},~\lambda_2=1-\mathcal{F}~.
\end{equation}
In our analysis, $\mathcal{F}\sim\mathcal{O}\left(\left(\frac{\sigma}{m}\right)^2\right)$. We can therefore obtain, the two entropies as
\begin{equation}\label{d.6}
\begin{split}
S\left(\rho_\Lambda^\text{diag}\right)&=-\left(\frac{1}{2}\ln\left(\frac{1}{2}+\frac{1}{2}\ln\left(\frac{1}{2}\right)\right)\right)\\
&=-\ln\left(\frac{1}{2}\right)\\
\implies S^\text{diag}&=\ln2
\end{split}
\end{equation}
and
\begin{equation}\label{d.7}
\begin{split}
S(\rho_\Lambda)&=-\mathcal{F}\ln\mathcal{F}-(1-\mathcal{F})\ln(1-\mathcal{F})\\
&\simeq-\mathcal{F}\ln\mathcal{F}-(1-\mathcal{F})(-\mathcal{F})\\
\implies S(\rho_\Lambda)&\simeq\mathcal{F}-\mathcal{F}\ln\mathcal{F}~.
\end{split}
\end{equation}
The relative entropy of coherence using eq.(\ref{d.1}) then reads
\begin{equation}\label{d.8}
\begin{split}
C_{\text{Rel}}(\rho_\Lambda)&=S\left(\rho^{\text{diag}}_\Lambda\right)-S(\rho_\Lambda)\\
&=\ln2+\mathcal{F}\ln\mathcal{F}-\mathcal{F}~.
\end{split}
\end{equation}
Substituting the analytical form of $\mathcal{F}$ in the above expression, we can write down the relative entropy of coherence as
\begin{widetext}
\begin{equation}\label{d.9}
C_{\text{Rel}}(\rho_\Lambda)=\ln2+\frac{2n+1}{8}\left(\frac{\cosh\alpha-1}{\cosh\alpha+1}\right)\left(\frac{\sigma}{m}\right)^2\ln\left[\frac{2n+1}{8}\left(\frac{\cosh\alpha-1}{\cosh\alpha+1}\right)\left(\frac{\sigma}{m}\right)^2\right]-\frac{2n+1}{8}\left(\frac{\cosh\alpha-1}{\cosh\alpha+1}\right)\left(\frac{\sigma}{m}\right)^2~.
\end{equation}
\end{widetext}
\section{Different measure of coherence for spin 1/2 particles under Lorentz boost in (3+1)-dimensions}
\label{4}
\noindent In this section we shall investigate spin-1/2 particles under Lorentz boost in (3+1)-dimensions. We consider that the particle can move in any arbitrary direction. In this scenario, the representation of the Wigner's little group takes the form
\begin{equation}\label{4d.1}
\begin{split}
\mathcal{D}(W(\Lambda,\mathbf{p})&=\frac{(p^0+m)\cosh\left(\frac{\alpha}{2}\right)+(\mathbf{p}.\hat{e}))\sinh\left(\frac{\alpha}{2}\right)}{\sqrt{(p^0+m)(p^0\cosh\alpha+(\mathbf{p}.\hat{e})\sinh \alpha+m)}}\\
&-\frac{i\sinh\left(\frac{\alpha}{2}\right) (\mathbf{\sigma}.(\mathbf{p}\times\hat{e})))}{\sqrt{(p^0+m)(p^0\cosh\alpha+(\mathbf{p}.\hat{e}))\sinh \alpha+m)}}.
\end{split}
\end{equation}
In our case the boost direction is given by $\hat{e}=\hat{z}$, as a result one can write down the Wigner's little group $\mathcal{D}(W(\Lambda,\mathbf{p})$ in a matrix form as
\begin{equation}\label{4d.2}
\begin{split}
\mathcal{D}(W(\Lambda,\mathbf{p}))=\frac{1}{\sqrt{\mathcal{A}\mathcal{B}}}
\begin{pmatrix}
\mathcal{C}&&\mathcal{E}-i\mathcal{F}\\
-(\mathcal{E}+i\mathcal{F})&&\mathcal{C}
\end{pmatrix}
\end{split}
\end{equation}
where $\mathcal{A},~\mathcal{B},~\mathcal{C},~\mathcal{E}$ and $\mathcal{F}$ are defined as
\begin{equation}\label{4d.3}
\begin{split}
\mathcal{A}&\equiv p^0+m,~\mathcal{B}\equiv p^0\cosh\alpha+p_z\sinh\alpha+m,\\
\mathcal{C}&\equiv(p^0+m)\cosh\left(\frac{\alpha}{2}\right)+p_z\sinh\left(\frac{\alpha}{2}\right),\\
\mathcal{E}&=p_x\sinh\left(\frac{\alpha}{2}\right),~\mathcal{F}=p_y\sinh\left(\frac{\alpha}{2}\right).
\end{split}
\end{equation}
Now the dispersion relation ($c=1$), takes the form ${p^0}^2=p^2+m^2$ where $p_x^2+p_y^2+p_z^2=p^2$. The determinant of the matrix given in eq.(\ref{4d.2}) takes the form
\begin{equation}\label{4d.4}
\det\left[\mathcal{D}(W(\Lambda,\mathbf{p}))\right]=\frac{\mathcal{C}^2+\mathcal{E}^2+\mathcal{F}^2}{\mathcal{A}\mathcal{B}}.
\end{equation}
Using eq.(\ref{4d.3}) it is easy to check that $\frac{\mathcal{C}^2+\mathcal{E}^2+\mathcal{F}^2}{\mathcal{A}\mathcal{B}}=1$ which is a consistency check for the fact that unitary matrices have determinant value equals to unity. Hence, from eq.(\ref{4d.4}), we obtain the relation
\begin{equation}\label{4d.5}
\mathcal{A}\mathcal{B}=\mathcal{C}^2+\mathcal{E}^2+\mathcal{F}^2~.
\end{equation}
The starting state of the system reads
\begin{equation}\label{4d.6}
|\psi\rangle=\sum_\sigma\int d^3\textbf{p}\psi(\textbf{p})|p\rangle\otimes c_\sigma|\sigma\rangle
\end{equation}
where $c_0=c_1=\frac{1}{\sqrt{2}}$ where the form of $\psi(\textbf{p})$ is given by
\begin{equation}\label{4d.6a}
\psi(\textbf{p})=\frac{1}{\sqrt{2\pi\sigma^{2n+3}\Gamma\left[n+\frac{3}{2}\right]}}\textbf{p}^ne^{-\frac{p^2}{2\sigma^2}}~.
\end{equation}
For the (3+1)-dimensional case our analysis significantly differs from the analysis in \cite{RiddhiASM} as for the spin part the state considered has maximum coherence in this work compared to a single $|0\rangle$ state in \cite{RiddhiASM}. The state in the boosted frame of reference then takes the form
\begin{equation}\label{4d.6b}
\begin{split}
|\psi_\Lambda\rangle=\frac{1}{\sqrt{2}}\sum_{\sigma,\sigma'=0}^1\int d^3\textbf{p} \psi(\textbf{p})\sqrt{\frac{(\Lambda p)^0}{p^0}}\mathcal{D}_{\sigma'\sigma}(W(\Lambda,\textbf{p}))|\Lambda \textbf{p},\sigma'\rangle.
\end{split}
\end{equation}
Making use of eq.(\ref{boostedrho}), we can write down the reduced density matrix as
\begin{widetext}
\begin{equation}\label{4d.7}
\begin{split}
\rho_\Lambda=&\text{Tr}_p[|\psi_\Lambda\rangle\langle\psi_\Lambda|]=\frac{1}{2}\int d^3\mathbf{p} |\psi(\textbf{p})|^2\biggr[\frac{(\mathcal{C}+\mathcal{E})^2+\mathcal{F}^2}{AB}|0\rangle\langle0|+\frac{\mathcal{C}^2-\mathcal{E}^2+\mathcal{F}^2+2i\mathcal{E}\mathcal{F}}{AB}|0\rangle\langle1|\\&+\frac{\mathcal{C}^2-\mathcal{E}^2+\mathcal{F}^2-2i\mathcal{E}\mathcal{F}}{AB}|1\rangle\langle0|+\frac{(\mathcal{C}-\mathcal{E})^2+\mathcal{F}^2}{AB}|1\rangle\langle1|\biggr]\\
=&\frac{1}{2}\int d^3\mathbf{p} |\psi(\textbf{p})|^2\biggr[\left(1+\frac{2\mathcal{C}\mathcal{E}}{AB}\right)|0\rangle\langle0|+\left(1-\frac{2\mathcal{E}(\mathcal{E}-i\mathcal{F})}{AB}\right)|0\rangle\langle1|+\left(1-\frac{2\mathcal{E}(\mathcal{E}+i\mathcal{F})}{AB}\right)|1\rangle\langle0|+\left(1-\frac{2\mathcal{C}\mathcal{E}}{AB}\right)|1\rangle\langle1|
\end{split}
\end{equation}
\end{widetext}
where in order to obtain the last line of the above equation, we have made use of eq.(\ref{4d.5}). Before, proceeding to calculate the individual elements of the reduced density matrix, we start by writing down the elements of momentum in spherical polar coordinates as $p_x=p\sin\theta\cos\phi,~p_y=p\sin\theta\sin\phi$, and $p_z=p\cos\theta$. As $p_z$ and $p_0$ are independent of $\phi$, the only dependence of the azimuthal angle comes via $\mathcal{E}$ and $\mathcal{F}$.

\noindent The elements of the density matrix then reads
\begin{align}
\rho_{\Lambda_{00}}&=\frac{1}{2}\int d^3\textbf{p} |\psi(\textbf{p})|^2\left(1+\frac{2\mathcal{C}\mathcal{E}}{AB}\right)~,\label{4d.8}\\
\rho_{\Lambda_{11}}&=\frac{1}{2}\int d^3\textbf{p} |\psi(\textbf{p})|^2\left(1-\frac{2\mathcal{C}\mathcal{E}}{AB}\right)~,\label{4d.9}\\
\rho_{\Lambda_{01}}&=\frac{1}{2}\int d^3\textbf{p} |\psi(\textbf{p})|^2\left(1-\frac{2\mathcal{E}(\mathcal{E}-i\mathcal{F})}{AB}\right)~,\label{4d.10}\\
\rho_{\Lambda_{10}}&=\frac{1}{2}\int d^3\textbf{p} |\psi(\textbf{p})|^2\left(1-\frac{2\mathcal{E}(\mathcal{E}+i\mathcal{F})}{AB}\right)~.\label{4d.11}
\end{align}

\noindent In order to obtain the elements we represent the three integrals into spherical polar coordinates. We start with eq.(\ref{4d.8}) as
\begin{equation}\label{4d.12}
\begin{split}
\rho_{\Lambda_{00}}=&\frac{1}{2}\int_0^\infty dp~p^2 |\psi(\textbf{p})|^2\int_0^{\pi} d\theta\sin\theta\int_0^{2\pi} d\phi \left(1+\frac{2\mathcal{C}\mathcal{E}}{AB}\right)\\
=&2\pi\int_0^\infty dp~p^2|\psi(\textbf{p})|^2+\frac{1}{2}\int_0^\infty dp~p^2 |\psi(\textbf{p})|^2\\&\times\int_0^\pi d\theta\sin\theta\left(\frac{2\mathcal{C}}{\mathcal{A}\mathcal{B}}\right) p\sin\theta\sinh\left(\frac{\alpha}{2}\right)\int_0^{2\pi} d\phi 
\cos\phi\\
=&2\pi\int_0^\infty dp~p^2|\psi(\textbf{p})|^2.
\end{split}
\end{equation} 
Substituting the form of $\psi(\textbf{p})$ from eq.(\ref{4d.6a}) and making a simple substitution $\xi_\sigma=\frac{p}{\sigma}$, we can recast the above equation as
\begin{equation}\label{4d.13}
\begin{split}
\rho_{\Lambda_{00}}=\frac{1}{\Gamma\left[n+\frac{3}{2}\right]}\int_0^\infty d\xi_\sigma \xi_\sigma^{2(n+1)}e^{-\xi_\sigma^2}~.
\end{split}
\end{equation}
The above integral converges for $n>-\frac{3}{2}$ and we obtain the form of $\rho_{\Lambda_{00}}$ as $\rho_{\Lambda_{00}}=\frac{1}{2}$. Similarly, one can solve the integral in eq.(\ref{4d.9}) to obtain $\rho_{\Lambda_{11}}$ as $\rho_{\Lambda_{11}}=\frac{1}{2}$.
\begin{widetext}
\noindent In order to obtain the off diagonal terms, we take the narrow-Gaussian width appoximation $\frac{\sigma}{m}\ll 1$ and simplify the $\left(1-\frac{2\mathcal{E}(\mathcal{E}-i\mathcal{F})}{AB}\right)$ term in eq.(\ref{4d.10}) as
\begin{equation}\label{4d.14}
\begin{split}
\left(1-\frac{2\mathcal{E}(\mathcal{E}-i\mathcal{F})}{AB}\right)&=1-2p_x\sinh\left(\frac{\alpha}{2}\right)\frac{p_x\sinh\left(\frac{\alpha}{2}\right)+p_y\sinh\left(\frac{\alpha}{2}\right)}{(p^0+m)(p^0\cosh\alpha+p_z\sinh\alpha+m)}\\
&\simeq1-\frac{\cosh\alpha-1}{\cosh\alpha+1}\left(\frac{p}{\sigma}\right)^2\left(\frac{\sigma}{m}\right)^2\sin^2\theta\left(\cos^2\phi-i\sin\phi\cos\phi\right).
\end{split}
\end{equation}
Similarly, the $\left(1-\frac{2\mathcal{E}(\mathcal{E}+i\mathcal{F})}{AB}\right)$ in this narrow-width approximation can be simplified up to $\mathcal{O}\left(\left(\frac{\sigma}{m}\right)^2\right)$ as
\begin{equation}\label{4d.15}
\left(1-\frac{2\mathcal{E}(\mathcal{E}+i\mathcal{F})}{AB}\right)\simeq 1-\frac{\cosh\alpha-1}{\cosh\alpha+1}\left(\frac{p}{\sigma}\right)^2\left(\frac{\sigma}{m}\right)^2\sin^2\theta\left(\cos^2\phi+i\sin\phi\cos\phi\right).
\end{equation}
Up to $\mathcal{O}\left(\frac{\sigma^2}{m^2}\right)$, the off diagonal elements of the reduced density matrix can then be obtained as
\begin{equation}\label{4d.16}
\begin{split}
\rho^\Lambda_{01}&=\frac{1}{2}\int _0^\infty dp\frac{p^{2(n+1)} e^{-\frac{p^2}{\sigma^2}}}{2\pi \sigma^{2n+3}\Gamma\left[n+\frac{3}{2}\right]}\int_0^\pi d\theta \sin\theta\int_0^{2\pi} d\phi \left(1-\frac{\cosh\alpha-1}{\cosh\alpha+1}\left(\frac{p}{\sigma}\right)^2\left(\frac{\sigma}{m}\right)^2\sin^2\theta\left(\cos^2\phi+i\sin\phi\cos\phi\right)\right)\\
&=\frac{1}{2}-\frac{\cosh\alpha-1}{4(\cosh\alpha+1)}\left(\frac{\sigma}{m}\right)^2\int_0^\infty d\left(\frac{p}{\sigma}\right)\left(\frac{p}{\sigma}\right)^{2(n+2)}\frac{e^{-\left(\frac{p}{\sigma}\right)^2}}{2\pi \Gamma\left[n+\frac{3}{2}\right]}\int_0^\pi d\theta\sin^3\theta\int_0^{2\pi} d\phi(\cos^2\phi
-i\sin\phi\cos\phi)\\
&=\frac{1}{2}-\frac{\cosh\alpha-1}{8\pi\Gamma\left[n+\frac{3}{2}\right](\cosh\alpha+1)}\left(\frac{\sigma}{m}\right)^2\int_0^\infty d\xi_\sigma\xi_\sigma^{2(n+2)}e^{-\xi_\sigma^2}\left(\frac{4}{3}\right)\left(\pi-0\right)\\
&=\frac{1}{2}-\frac{2n+3}{24}\left(\frac{\sigma}{m}\right)^2\left(\frac{\cosh\alpha-1}{\cosh\alpha+1}\right)
\end{split}
\end{equation}
\end{widetext}
and 
\begin{equation}\label{4d.17}
\rho_{\Lambda_{10}}=\frac{1}{2}-\frac{2n+3}{24}\left(\frac{\sigma}{m}\right)^2\left(\frac{\cosh\alpha-1}{\cosh\alpha+1}\right).
\end{equation}
Again, defining a new quantuty $\mathcal{R}$ as
\begin{equation}\label{4d.18}
\mathcal{R}\equiv\frac{2n+3}{24}\left(\frac{\sigma}{m}\right)^2\left(\frac{\cosh\alpha-1}{\cosh\alpha+1}\right)
\end{equation}
we can express the reduced density matrix as
\begin{equation}\label{4d.19}
\rho_\Lambda=
\begin{pmatrix}
\frac{1}{2}&&\frac{1}{2}-\mathcal{R}\\
\frac{1}{2}-\mathcal{R}&&\frac{1}{2}
\end{pmatrix}~.
\end{equation}
From the above form of the reduced density matrix with real elements, it is easy to check that $\rho_{01}=\rho_{10}$ with the diagonal elements being given by $\frac{1}{2}$. Hence, we expent that the Frobenius-norm measure of coherence will same as the $l_1$-norm measure of coherence. We start by calculating th two coherence measures for the reduced desity matrix of the system. The eigenvalues of the reduced density matrix read
\begin{equation}\label{4d.20}
\lambda_1=\mathcal{R},~\lambda_2=1-\mathcal{R}.
\end{equation}
The Frobenius-norm measure of coherence reads
\begin{equation}\label{4d.21}
\begin{split}
C_F(\rho_\Lambda)&=\sqrt{2}\sqrt{\left(\lambda_1-\frac{1}{2}\right)^2+\left(\lambda_2-\frac{1}{2}\right)^2}\\
&=2\left(\frac{1}{2}-\mathcal{R}\right)\\
\implies C_F(\rho_\Lambda)&=1-2\mathcal{R}\\&=1-\frac{2n+3}{12}\left(\frac{\sigma}{m}\right)^2\left(\frac{\cosh\alpha-1}{\cosh\alpha+1}\right).
\end{split}
\end{equation}
A few comments are in order now. We note that for $n=0$, the above result for the $(3+1)$-dimensional case is exactly identical to the $(1+1)$-dimensional case given in eq.(\ref{l1.2}). The two results differ from each other when $n$ is greater than zero. One can also find a bound on $n$ from the condition that $C_F(\rho_\Lambda)\geq0$ which leads to the result (in the $v\rightarrow c$ limit)
\begin{equation}\label{4d.21a}
n\leq \left[6\left(\frac{m}{\sigma}\right)^2-\frac{3}{2}\right].
\end{equation}
Again the Frobenius-norm measure of coherence must be less than or equal to unity. In the boosted frame of reference the coherence measure should be less than unity (coherence measure unity refers to complete coherence) which from eq.(\ref{4d.21}), gives a lower bound to $n$ which is
\begin{equation}\label{4d.21b}
n>-\frac{3}{2}~.
\end{equation}
From the discussion after eq.(\ref{4d.13}), we find that the integral converges only for $n>-\frac{3}{2}$ which is exactly similar to the bound obtained using the physical arguments for the values of the coherence measure.
Combining the above two inequalities, we obtain a range for $n$ as
\begin{equation}\label{4d.21c}
-\frac{3}{2}<n\leq\left[6\left(\frac{m}{\sigma}\right)^2-\frac{3}{2}\right]~.
\end{equation}
One can make an important observation by comparing the above result with the range of $n$ in eq.(\ref{n_range}) for the (1+1)-dimensional case. It can be observed the upper and lower bounds for $n$ has exactly trippled for the (3+1)-dimensional case in comparison to the bounds obtained in the (1+1)-dimesnional case. This result can be a direct consequence of the degrees of freedom getting trippled corresponding to each spin-1/2 particle in the (3+1)-dimensional case, compared to the (1+1)-dimensional case.

\noindent Now, we proceed to calculate the $l_1$-norm measure of coherence for this (3+1)-dimensional case and compare it with the Frobenius norm-measure of coherence. The $l_1$-norm measure of coherence reads
\begin{equation}\label{4d.22}
\begin{split}
C_{l_1}(\rho_\Lambda)&=|\rho_{\Lambda_{01}}|+|\rho_{\Lambda_{10}}|\\
&=2|\rho_{\Lambda_{01}}|\\
\implies C_{l_1}(\rho_\Lambda)&=1-2\mathcal{R}~.
\end{split}
\end{equation}
Comparing eq.(\ref{4d.21}) with eq.(\ref{4d.22}), we can find (as expected) the Frobenius-norm measure of coherence is exactly equal to the $l_1$-norm measure of coherence. The $l_2$ norm measure of coherence reads
\begin{equation}\label{4d.23}
\begin{split}
C_{l_2}(\rho_\Lambda)&=|\rho_{\Lambda_{01}}|^2+|\rho_{\Lambda_{10}}|^2\\&\simeq\frac{1}{2}-2\mathcal{R}\\
&=\left(\frac{1}{2}-\mathcal{R}\right)^2+\left(\frac{1}{2}-\mathcal{R}\right)^2\\
&=\frac{1}{2}-\frac{2n+3}{12}\left(\frac{\sigma}{m}\right)^2\left(\frac{\cosh\alpha-1}{\cosh\alpha+1}\right).
\end{split}
\end{equation}
One can now obtain the relative entropy of coherence using eq.(\ref{d.1}) and following the procedure in subsection (\ref{3.d}) as (up to $\mathcal{O}\left(\left(\frac{\sigma}{m}\right)^2\right)$)
\begin{equation}\label{4d.24}
\begin{split}
C_{\text{Rel}}(\rho_\Lambda)&=S(\rho^\text{diag}_\lambda)-S(\rho_\Lambda)\\
&=\ln2+\mathcal{R}\ln\mathcal{R}+(1-\mathcal{R})\ln(1-\mathcal{R})\\
\implies C_{\text{Rel}}(\rho_\Lambda)&\simeq\ln2+\mathcal{R}\ln\mathcal{R}-\mathcal{R}
\end{split}
\end{equation}
where the reduced density matrix $\rho_\Lambda$ is given in eq.(\ref{4d.19}). Substituting the analytical form of $\mathcal{R}$, we obtain the relative entropy of coherence for the  system as
\begin{widetext}
\begin{equation}\label{4d.25}
C_{\text{Rel}}(\rho_\Lambda)=\ln2+\frac{2n+3}{24}\left(\frac{\cosh\alpha-1}{\cosh\alpha+1}\right)\left(\frac{\sigma}{m}\right)^2\ln\left[\frac{2n+3}{24}\left(\frac{\cosh\alpha-1}{\cosh\alpha+1}\right)\left(\frac{\sigma}{m}\right)^2\right]-\frac{2n+3}{24}\left(\frac{\cosh\alpha-1}{\cosh\alpha+1}\right)\left(\frac{\sigma}{m}\right)^2~.
\end{equation}
\end{widetext}
\section{Different measures of coherence and their corresponding plots}
\label{5}
\noindent In this section, we shall plot the Frobenius-norm measure of coherence that we have obtained in the earlier section with respect to the change in Gaussian width for different values of the boost parameter.
\subsection{Plot of $\mathbf{C_{l_1}(\rho_\Lambda)}$ (or $C_F(\rho_\Lambda)$) vs $\sigma$}
\noindent At first we are going to consider the coherence measure obtained in eq.(\ref{co_me(1+1)}) for electron with rest mass $m\approx 0.5 \text{ MeV}$. For this case, the measure of coherence is valid upto $\sigma<0.5 \text{ MeV}$. In eq.(\ref{co_me(1+1)}), $\cosh\alpha=\frac{1}{\sqrt{1-\beta^2}}$ where $\beta=\frac{v}{c}$ is the boost parameter. Now the Frobenius-norm measure of coherence as has been observed in subsection (\ref{3.a}) is exactly same up to $\mathcal{O}\left(\left(\frac{\sigma}{m}\right)^2\right)$ to the $l_1$-norm measure of coherence obtained in subsection (\ref{3.b}). 
\begin{figure}[ht!]
\includegraphics[scale=0.57]{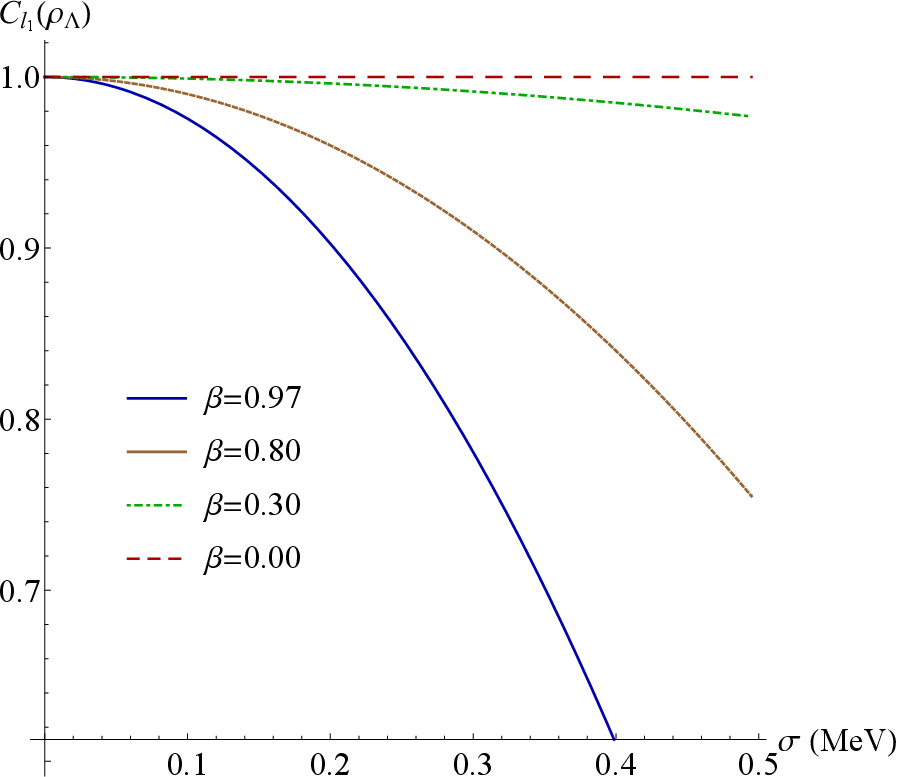}
\caption{The $l_1$ norm measure of coherence, $C_{l_1}(\rho_\Lambda)$ vs $\sigma$ (for electron) with the values of the boost parameter $\beta~=~0.97,~0.8,~0.3,~0.0$ and $n=1.5$.}
\label{F1}
\end{figure}
As $l_1$-norm measure of coherence is often considered to be more accurate coherence monotone, we plot this against $\sigma$. The plot of $C_{l_1}(\rho)$ (for electrons) vs $\sigma$ for different values of $\beta$ and a fixed value of $n=1.5$ is given in Figure(\ref{F1}). From Figure(\ref{F1}), we observe that there is a substantial amount of loss in coherence for large boosts (e.g. $\beta=0.80,0.97$) with the change in the parameter $\sigma$ in contrast to lower values of the boost parameter(e.g. $\beta=0.3,0.0$) for an electron.
\begin{figure}[ht!]
\includegraphics[scale=0.57]{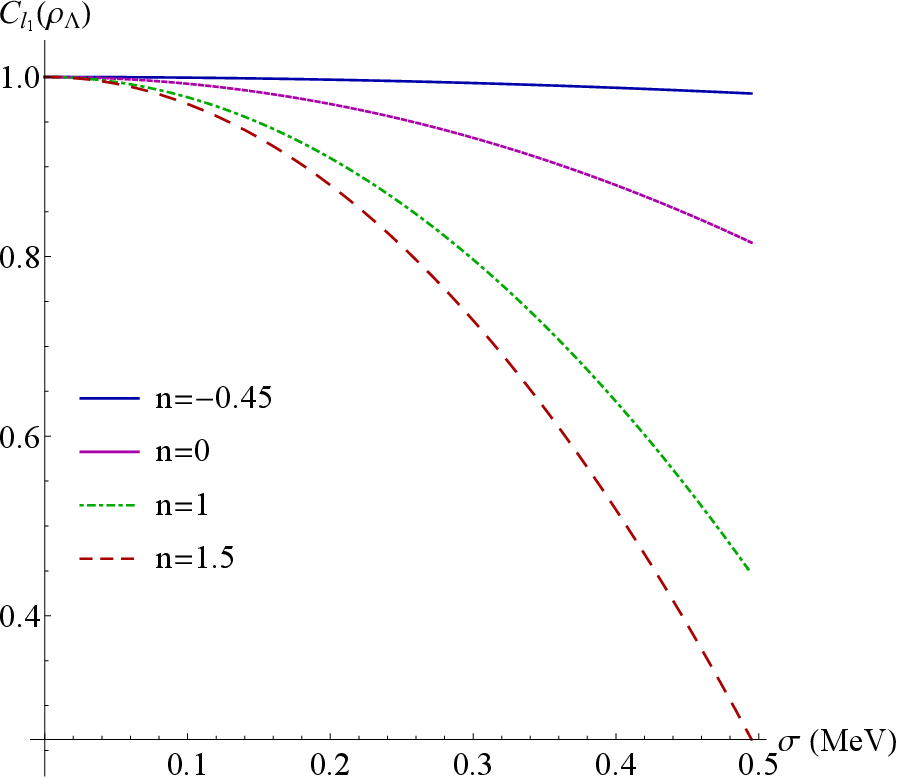}
\caption{$C_{l_1}(\rho)$ vs $\sigma$ (for electron) with  $n~=~-0.45,~0,~1,~1.5$ and $\beta=0.99$.}
\label{F2}
\end{figure}
The coherence is being measured between the momentum and spin state of the electron. The boost enhances the momentum parameter corresponding to a loss of coherence between the two states. We shall next investigate the dependence of the coherence-measure for a fixed value of the boost parameter with different values of $n$. We start by considering $\beta=0.99$ and take values of $n$ which lies within the range of $n$ obtained in eq.(\ref{n_range}). As $n>-\frac{1}{2}$, we take the minimum value of $n$ to be $-0.45$ whereas the maximum value is taken to be same as the minimum upper bound of $n$ that is $1.5$ in case of the electron. We consider the (1+1)-dimensional model corresponding to the electron. From this subsection it is evident that the loss of coherence becomes significant with the increase in the boost parameter $\beta$ and $n$. From Fig.(\ref{F2}), it can be seen that with increasing value of $n$ the loss of coherence is higher when $C_{l_1}(\rho_\Lambda)$ is plotted against $\sigma$.
\begin{figure}[ht!]
\includegraphics[scale=0.57]{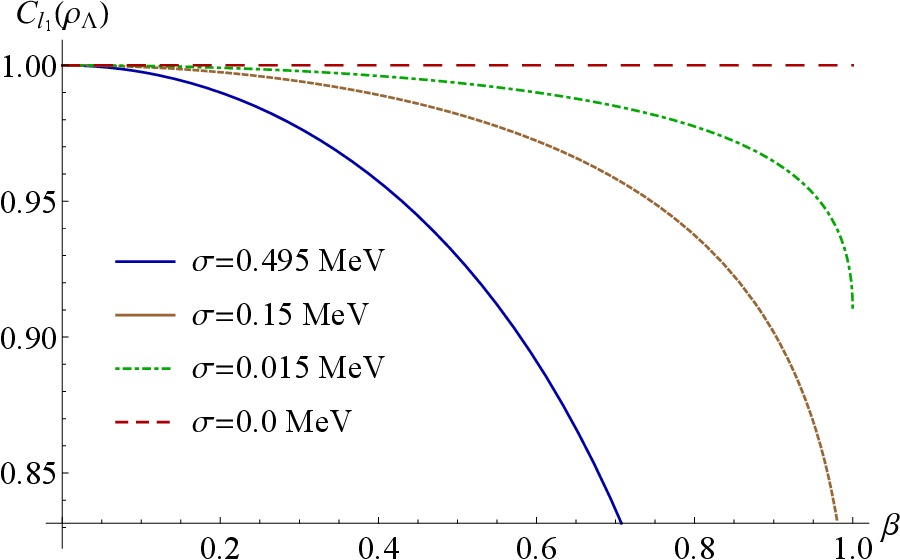}
\caption{Plot of $C_{l_1}(\rho)$ vs $\beta$ (for electron) with $\sigma=0.495 \text{ MeV},~0.15 \text{ MeV},~0.015 \text{ MeV},~0.0$ and $n=1.5$.}
\label{F3}
\end{figure}
We next plot $C_{l_1}(\rho_\Lambda)$ vs $\beta$ with fixed value of $n=4.5$, and different values of the $\sigma$ parameter in Fig.(\ref{F3}). It is important to note that even with an increase in the boost parameter the coherence loss is non-existent when the Gaussian-width corresponding to the momentum represent of the electron wave-function is zero. This kind of scenario represents a spiked wave-packet resembling the Dirac-delta function (which is quite evident in the $\sigma\rightarrow0$ limit for the $n=0$ case in eq.(\ref{fx})). Hence, the more de-localized is the momentum wave function, the loss of coherence due to boost is more significant.

\noindent We next consider the coherence measure for the case of a neutron. For neutron, we consider the (3+1)-dimensional model developed in section (\ref{4}). From the forms of the Frobenius-norm measure of coherence in eq.(\ref{4d.21}) and the $l_1$-norm measure of coherence in eq.(\ref{4d.22}), it is evident that the two measures of coherence are identical to each other in the (3+1)-dimensional case as well. 
\begin{figure}[ht!]
\includegraphics[scale=0.57]{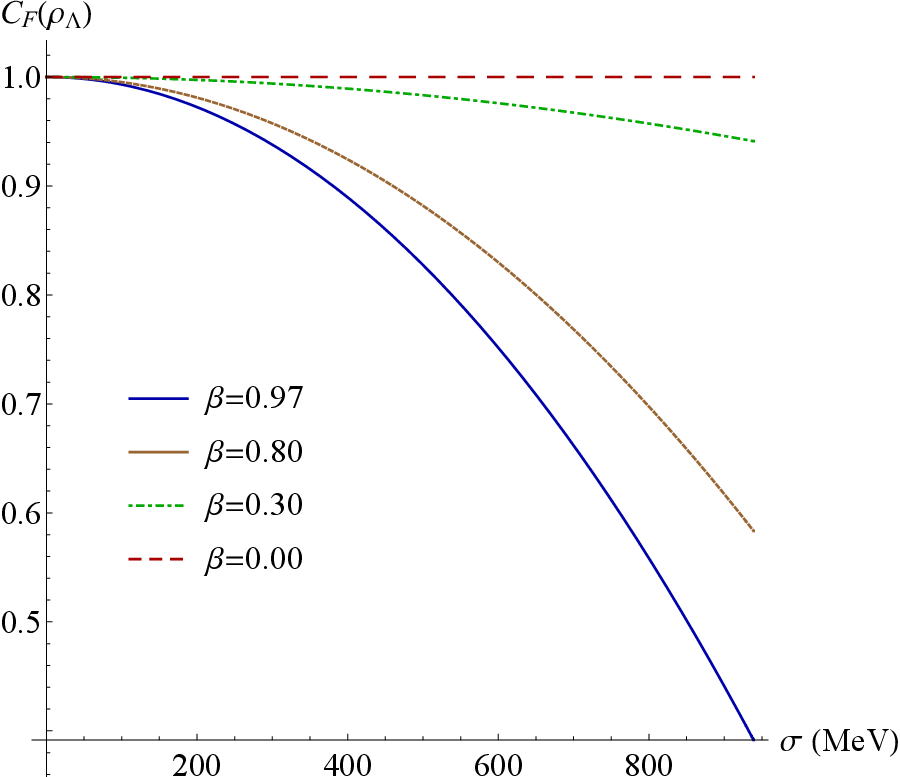}
\caption{$C_F(\rho)$ vs $\sigma$ (for neutron) with $\beta~=~0.97,~0.8,~0.3,~0.0$ and $n=1.5$.}
\label{F4}
\end{figure}
As a result, we plot the the Frobenius-norm measure of coherence for a neutron with a rest mass energy $m=939.36 \text{ MeV}$. As before we have considered $\sigma<939.36\text{ MeV}$ and we have plotted $C_F(\rho_\Lambda)$ up to $\sigma=939~\text{MeV}$ in Figure (\ref{F4}). Due to the heavier rest mass energy of the neutron we observe a slower loss of coherence in neutron.
\begin{figure}[ht!]
\includegraphics[scale=0.57]{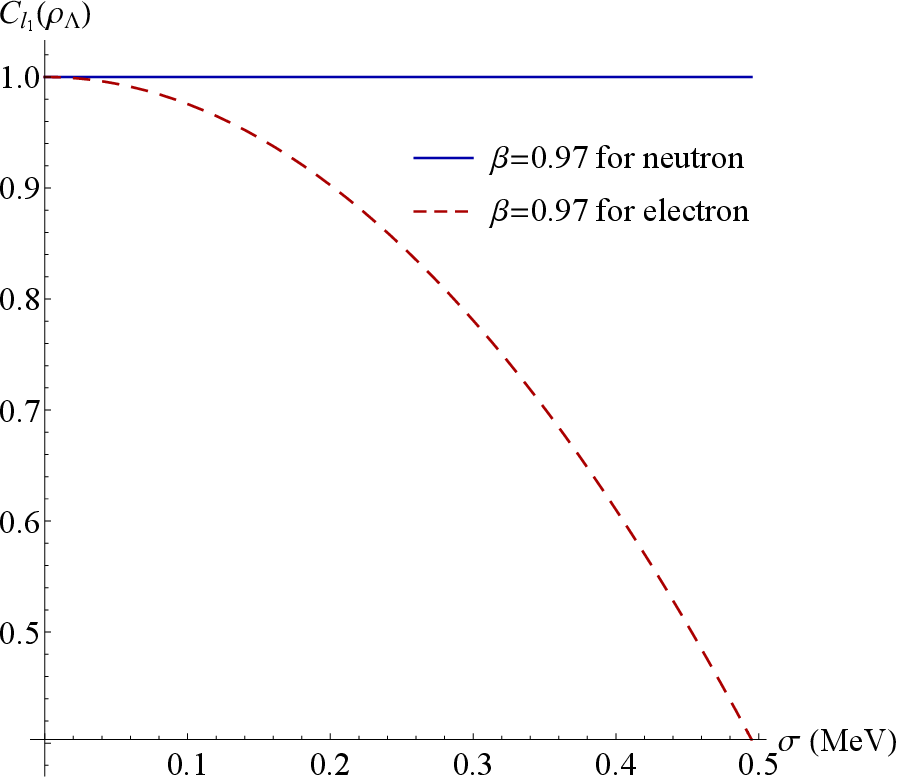}
\caption{$C_F(\rho)$ vs $\sigma$ with $\beta~=~0.97 $ and $n=1.5$ for electron and neutron.}
\label{F5}
\end{figure}
Finally, we plot the loss of coherence corresponding to the neutron with respect to the electron. To do so, we start by restricting the width of the Gaussian wave-packet up to the value of the lower rest mass energy which is of the electron. In Fig.(\ref{F5}), we observe that for increase in $\sigma$ the loss of coherence is significant in the electron when the value of the boost parameter is set to $\beta=0.97$. This implies that due to the heavier rest mass energy of the neutron coherence loss for a fixed wave-packet width is significantly smaller. In the next, two subsections, we shall investigate the dependence of the $l_2$-norm measure of coherence and relative entropy of coherence with the boost parameter. Hence, we observe similar dependence of the coherence measure on the boos parameter both for an electron and neutron compared to the $l_1$-norm (or Frobenius norm) case.
%\noindent In Figure(\ref{F4}), we plot $C_F(\rho)$ for both electron and neutron ($n=2$) and with the boost parameter fixed at $\beta=0.99$. We have plotted upto $\sigma=0.49\text{ MeV}$ as beyond $\sigma=0.5\text{ MeV}$ the analytical form of $C_F(\rho)$ for the electron does not hold any more (as it violates the $\sigma<m~ (0.5 \text{ MeV})$ condition). We observe a considerable amount of loss in the measure of coherence for an electron (than the neutron) even for a high boost parameter ($\beta=0.99$). Next, we shall look at the dependence of the Frobenius norm measure of coherence on the parameter $n$ in case of an electron. With $\beta=0.99$, $\sigma=0.49\text{ MeV}$ and $m=0.5MeV$ (along with the condition $C_F(\rho)\geq 0$), we get $n<2.27 $ from the relation in eq.(\ref{co_me(1+1)}). Therefore, we plot $C_F(\rho)$ vs $\sigma$ for $n=0,~1$ and $2$ in Figure(\ref{F5}). We observe from Figure(\ref{F5}) that with the increase in the parameter $n$, the loss of coherence becomes more significant than before.
\textcolor{blue}{\begin{figure}[ht!]
\includegraphics[scale=0.57]{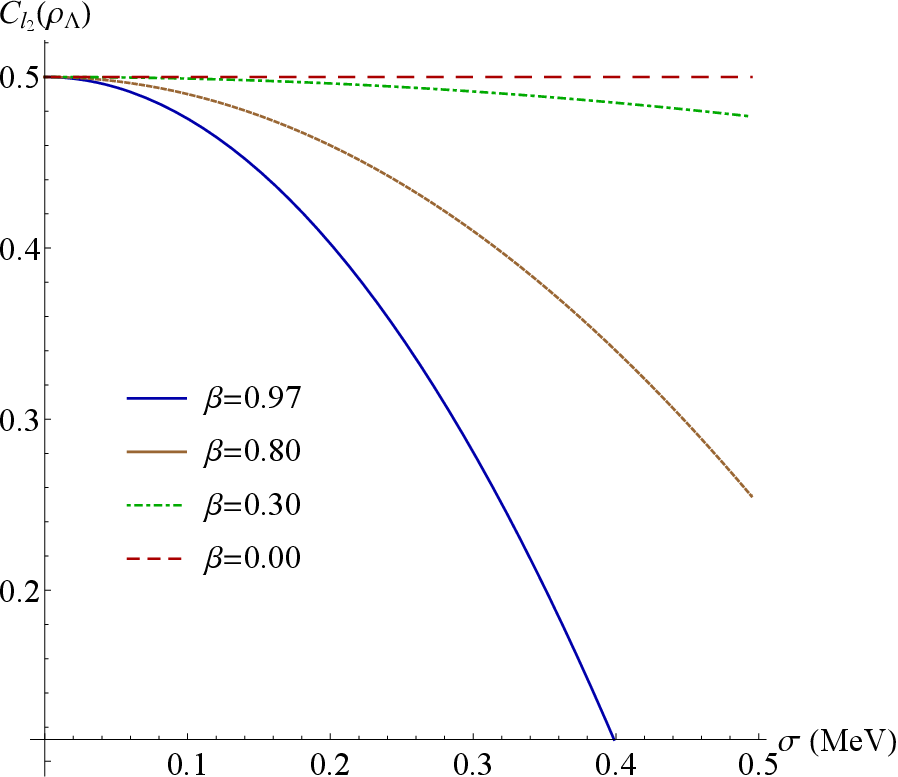}
\caption{$C_{l_2}(\rho_\Lambda)$ vs $\sigma$ (for electron) with  $n~=~1.5$ and $\beta~=~0.97,~0.8,~0.3,~0.0$.}
\label{F6}
\end{figure}}
\subsection{$\mathbf{C_{l_2}(\rho_\Lambda)}$ vs $\mathbf{\sigma}$ plot}
\noindent In this subsection, we shall plot the $l_2$ norm measure of coherence for an electron as well as neutron against $\sigma$ for different values of the boost parameter.

\noindent For the electron case, we use the (1+1)-dimensional result in eq.(\ref{l2.2}) and set $n=1.5$. We have then plotted $C_{l_2}(\rho_\Lambda)$ vs $\sigma$ for different values of the boost parameter ($\beta=0.97,~0.8,~0.3,~0$) in Fig.(\ref{F6}).
\noindent As observed in the earlier plots, we also observe an  increase in the loss of coherence for increasing values of the boost parameter. 
\begin{figure}[ht!]
\includegraphics[scale=0.57]{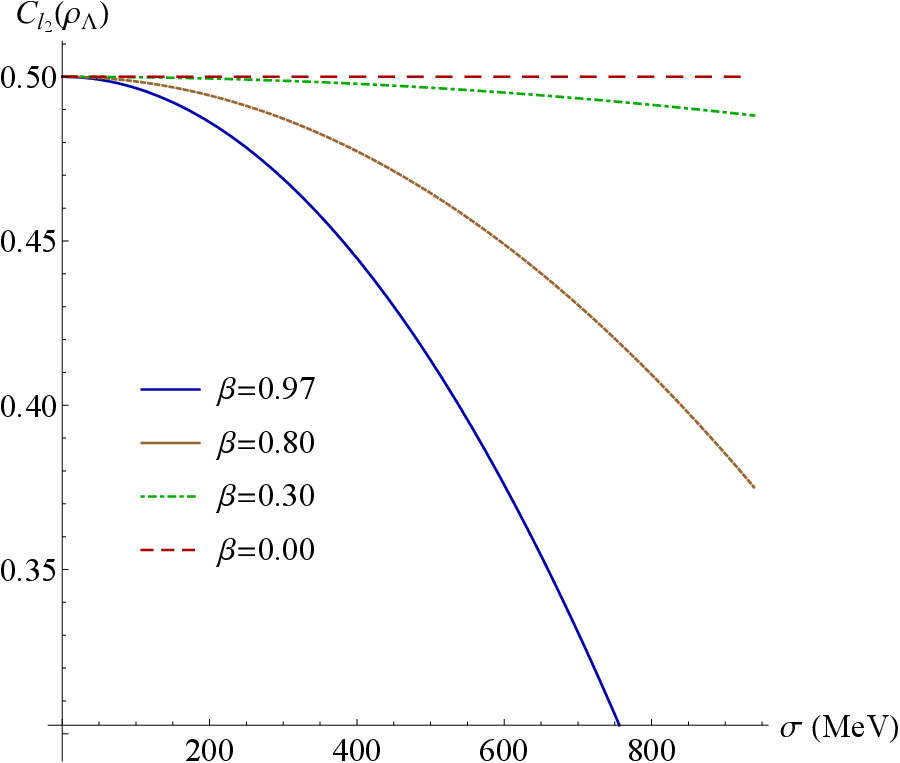}
\caption{$C_{l_2}(\rho_\Lambda)$ vs $\sigma$ (for neutron) with  $n~=~4.5$ and $\beta~=~0.97,~0.8,~0.3,~0.0$.}
\label{F7}
\end{figure}
Next, we consider the case of the neutron and make use of the analytical form of $C_{l_2}(\rho_\Lambda)$ given in eq.(\ref{4d.23}) for a fixed value of $n=4.5$. We plot $C_{l_2}(\rho_\Lambda)$ vs $\sigma$ for different values of the boost parameter in Fig.(\ref{F7}). We also find out that for zero boost the coherence loss is zero and it becomes more and more significant for higher values of the parameter $\beta$.
\subsection{$C_{\text{Rel}}(\rho_\Lambda)$ vs $\sigma$ plot}
\begin{figure}[ht!]
\includegraphics[scale=0.57]{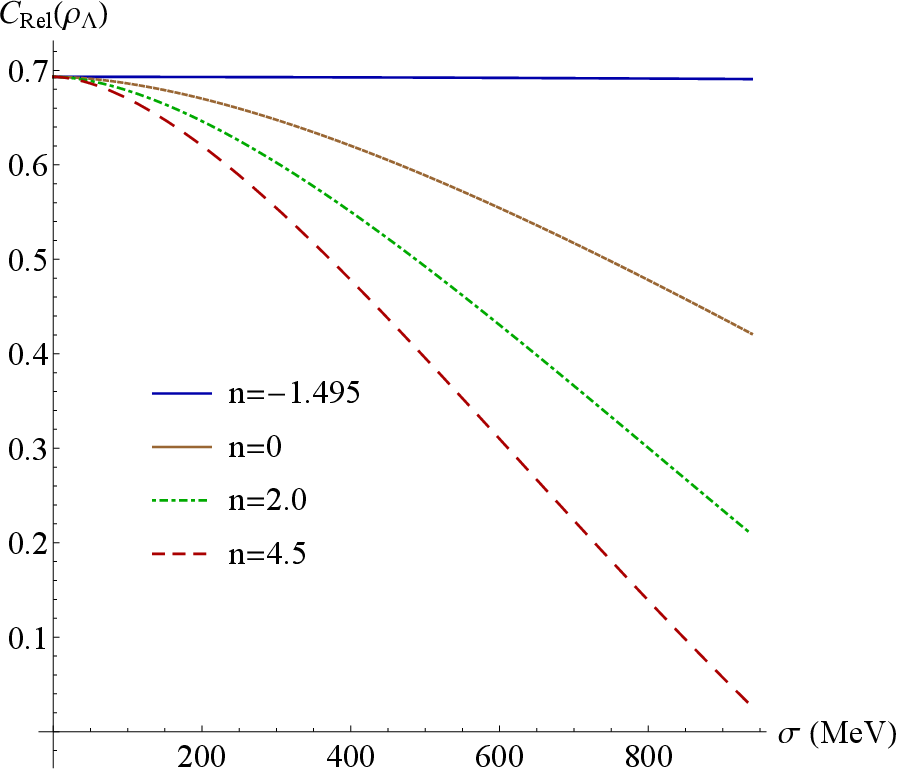}
\caption{$C_{\text{Rel}}(\rho_\Lambda)$ vs $\sigma$ (for neutron) with  $\beta=0.97$ and $n~=~-1.495,~0,~2,~4.5$.}
\label{F10}
\end{figure}

\noindent Finally, we consider the relative measure of coherence for the cases described earlier. We start by plotting $C_{\text{Rel}}(\rho_\Lambda)$ against the width of the generalized Gaussian momentum wave-function for fixed values of the boost-parameter.
\begin{figure}[ht!]
\includegraphics[scale=0.57]{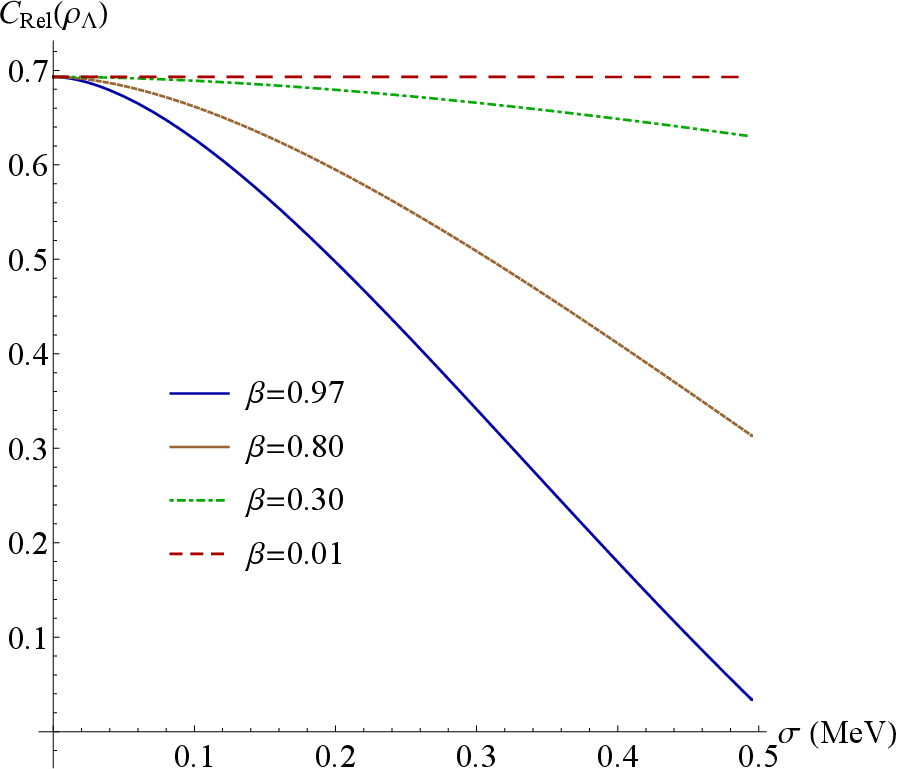}
\caption{$C_{\text{Rel}}(\rho_\Lambda)$ vs $\sigma$ (for electron) with  $n~=~1.5$ and $\beta~=~0.97,~0.8,~0.3,~0.01$.}
\label{F8}
\end{figure}
As the leading term in the relative measure of coherence both in  eq.(s)(\ref{d.9},\ref{4d.25}) is $\ln 2\simeq0.693$, the complete value of coherence can not exceed $ln 2$ unlike the upper-bound 1 in the $l_1$-norm or Frobenius norm case. For electron we make use of eq.(\ref{d.9}) and plot against $\sigma$ for $n=1.5$ with different values of $\beta$. We find the same behaviour of $C_{\text{Rel}}(\rho_\Lambda)$ with $\beta$ in Fig.(\ref{F8}), when an electron is being considered. We observe that there is minimal loss of coherence for a very small value of the boost parameter ($\beta=0.01$ in the plot) with increase in the Gaussian-width.
\begin{figure}[ht!]
\includegraphics[scale=0.57]{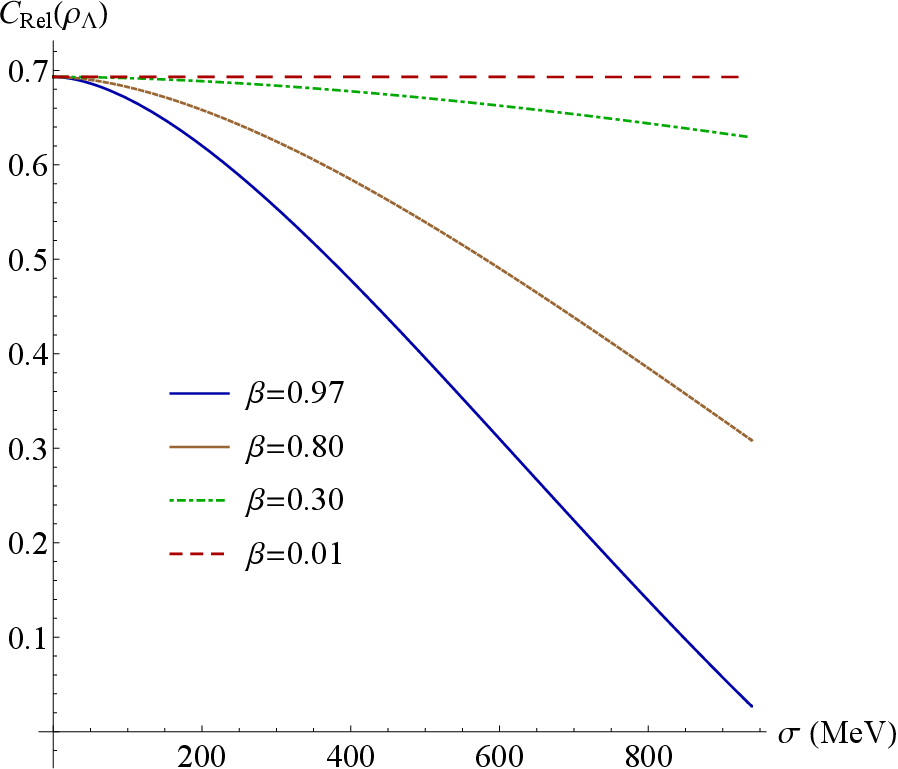}
\caption{$C_{\text{Rel}}(\rho_\Lambda)$ vs $\sigma$ (for neutron) with  $n~=~4.5$ and $\beta~=~0.97,~0.8,~0.3,~0.01$.}
\label{F9}
\end{figure}
\noindent We can also observe the same behaviour for the neutron case where we have made use of eq.(\ref{4d.25}) and the value of $n$ is fixed to 4.5 in Fig.(\ref{F9}). We find out that with increase in $\beta$ the coherence loss is much more significant. To investigate the dependence of the $n$-parameter on the coherence loss among the momentum and spin part of a neutron we have further plotted $C_{\text{Rel}}(\rho_\Lambda)$ against $\sigma$ with a fixed value of $\beta$ ($0.97$) in Fig.(\ref{F10}). We observe that  near the lower-bound for $n$ as has been obtained in eq.(\ref{4d.21c}), the coherence loss becomes negligible for increase in the parameter $\sigma$. This implies that the coherence-loss can be made minimal with a generalized wave packet where the value of $n$ is very close to the lower-bound for $n$.

\section{Discussion and Conclusion}\label{6}
\noindent In this work, we consider a single particle entangled state and its quantum coherence under Lorentz boost. At first, we considered a generalized form of the Gaussian wave packet (with its peak at the origin), where the general Gaussian wave function (in momentum space) is multiplied by $n$-th power of the momentum $p$. For a given boost in one direction ($(1+1)$-dimensional consideration) we have obtained different measures of coherence with a dependence on the boost parameter $\beta$, Gaussian width $\sigma$, the mass of the fermion, and the generalization parameter $n$. At first, we have calculated the Frobenius-norm measure  of coherence and then we have calculated the $l_1$ norm measure of coherence for the system. We observe that $C_F(\rho_\Lambda)$ comes out to be exactly similar to $C_{l_1}(\rho_\Lambda)$. We find out that such a coincidence can occur if and only if the diagonal elements of the reduced density matrix are $\frac{1}{2}$ and the off-diagonal elements are complex-conjugate of each other. In our case the off-diagonal elements are real and equal to each other which results in such a beautiful coincidence. Next, we make use of the condition that these two measures of coherence neither can be greater than unity nor can be less than zero in the $\beta\rightarrow1$ limit. Implementation of these two condition provided us with a bound on the ``generalization parameter '' $n$ with $0.5$ being the lower-bound and $2\left(\frac{\sigma}{m}\right)^2-\frac{1}{2}$ in the (1+1) dimensional scenario. Next, we have investigated the $l_2$ norm measure of coherence and relative entropy of coherence for the above case in consideration. In the next section (section (\ref{4})), we have used a $(3+1)$-dimensional model and investigated the similar measures of coherence. Comparing the form of the reduced density matrix in eq.(\ref{4d.19}) with eq.(\ref{rhomat}) we observe that for $n=0$, the two density matrices are identical which is a very important finding in our work. The reduced density matrix is again obtained with equal and real off-diagonal terms and with diagonal terms equal to $\frac{1}{2}$. Hence, according to the equality condition of the Frobenius-norm  measure of coherence and $l_1$-norm measure of coherence obtained for the (1+1)-d case, this reduced-density matrix again lead us to same analytical forms corresponding to the two measures of coherence. Again claiming that the coherence-measures has to be less than or equal to 1 and greater than or equal to zero, we obtain a range for the $n$-parameter. We find out from eq.(\ref{4d.21c}), that the lower bound for $n$ is -1.5 whereas the minimum value of the upper bound is equal to $6\left(\frac{\sigma}{m}\right)^2-\frac{3}{2}$. It is interesting to see that both of the bounds for $n$ for the (3+1)-dimensional case is exactly three-times to the bounds of $n$ in the (1+1)-dimensional scenario. The physical reason behind this phenomenon is that the degrees of freedom corresponding to each particle (while considering the spatial components) has exactly tripled in the (3+1)-dimensional scenario than the (1+1)-scenario. We have then obtained the analytical form of the $l_2$ norm of coherence and relative entropy of coherence. Finally, in section (\ref{5}), we plot different measures of coherence against the width of the Gaussian wave-packet. Primarily, we have considered the results obtained in section (\ref{3}) for the electron and the results in section (4) for the case of neutron. From Figure(\ref{F1}) in section(\ref{5}), we observe that for a zero Lorentz boost there will be no loss in the Frobenius-norm measure of coherence for an electron. We also observe that for a relativistic observer there will be a significant amount of loss in the measure of coherence depending on the value of the Lorentz boost where the loss becomes more and more significant for increase in $\beta$. Due to the Wigner rotation \cite{Weinberg} of a single-particle entangled state under Lorentz boost, the spin and momentum degrees of freedom couples with each other which results in loss of coherence when boost in non-zero. We also plot this measure of coherence with respect to the Gaussian width parameter for a fixed value of Lorentz boost and different values of the generalization parameter $n$ in the case of an electron in Fig.(\ref{F2}). We observe that with the increase in the parameter $n$, the loss in the measure of coherence becomes much more significant than before and for a value of $n$ close to the lower bound obtained in eq.(\ref{d.9}), we observe almost a negligible amount of coherence loss even for a very large value of the boost parameter when $C_{l_1}(\rho_\Lambda)$ is plotted against $\sigma$. Next, we have plotted $C_{l_1}(\rho_\Lambda)$ against $\beta$ for fixed values of the  Gaussian width ($\sigma$) and the generalization parameter $n$ in Fig.(\ref{F3}). We, observe that the coherence loss becomes significant when the wave-packet is de-localized. For a delta-function like wave packet, where the Gaussian width is zero, the loss of coherence is negligible even with $\sigma\sim m$ and $n=1.5$ (which is the upper-bound obtained for $n$ considering $\sigma\sim m$). For the next plot, we have considered a single particle along with a Lorentz boost in all three spatial directions. In this case, our main motivation is to consider a neutron as a single particle. We then also observe a similar loss in the measure of coherence from the perspective of a relativistic observer and observe significant coherence-losses with increasing value of the boost parameter in Fig.(\ref{F4}). In Fig.(\ref{F5}), we observe that if the width of the Gaussian wave-packet is fixed to a value such that $\sigma\sim m_{\text{electron}}$, then the loss in the measure of coherence for the neutron becomes negligible compared to the electron (as then $\frac{\sigma}{m_{\text{neutron}}}\sim \frac{m_{\text{electron}}}{m_\text{neutron}}\ll1$). This happens because of the very high rest mass energy of the neutron which is almost 1879 times higher than the electron. Next, we have plotted the $l_2$-norm of coherence and finally the relative entropy of coherence for the electron as well as neutron which also indicates similar behaviour of coherence on the boost parameter, Gaussian-width, and the generalization parameter as observed in the $l_1$ or Frobenius-norm case. Our analysis amplifies on the fact that both basis dependent and independent measure of coherence plays identical role in order to use coherence as a resource in relativistic framework which is opposite to the observation made in \cite{RiddhiASM}. Unlike \cite{RiddhiASM}, we find out that the loss-of coherence in case of a neutron is not negligible but only negligible if the Gaussian-width is made to be restricted to a very small regime such that $\sigma\ll m_{\text{neutron}}$. When $\sigma\sim m_{\text{neutron}}$, the loss of coherence even for the neutron is quite significant. This observation is in complete contrast to the one made in \cite{RiddhiASM}.
\section*{Acknowledgement}
\noindent We thank the anonymous referees for detailed review on our paper which has helped us to substantially improve our manuscript.
\appendix
\section{Single particle quantum state under Lorentz boost}
\noindent In this Appendix, we shall derive the form of the single particle quantum state after Lorentz boost has been applied\footnote{We shall essentially follow the analysis in \cite{Weinberg}.}.
Under a pure Lorentz boost $\Lambda$,  $\mathcal{U}(\Lambda,a)$ reads
\begin{eqnarray}
\mathcal{U}(\Lambda,a)\vert \mathbf{p},\sigma \rangle \rightarrow \mathcal{U}(\Lambda,0)\vert \mathbf{p},\sigma \rangle \equiv \mathcal{U}(\Lambda)\vert \mathbf{p},\sigma \rangle\,.
\end{eqnarray}
\\
New eigenvalues of the momentum operator $P^{\mu}$ become
\begin{eqnarray}
P^{\mu}\big[\mathcal{U}(\Lambda)\vert \mathbf{p},\sigma \rangle \big]&=&\left(\mathcal{U}(\Lambda)\,\mathcal{U}^{-1}(\Lambda)\right)P^{\mu}\big[\mathcal{U}(\Lambda)\vert \mathbf{p},\sigma \rangle \big]\nonumber\\
&=& \mathcal{U}(\Lambda)\big[\mathcal{U}^{-1}(\Lambda)P^{\mu}\mathcal{U}(\Lambda)\big]\vert \mathbf{p},\sigma \rangle \nonumber\\
&=& \mathcal{U}(\Lambda)\big[\Lambda^{\mu}_{\hspace{0.2cm}\nu}\,P^{\nu}\big]\vert \mathbf{p},\sigma \rangle \nonumber\\
&=&\Lambda^{\mu}_{\hspace{0.2cm}\nu}\,p^{\nu}\big[\mathcal{U}(\Lambda)\vert \mathbf{p},\sigma \rangle \big]~.
\end{eqnarray}
To derive the previous equation, we have used the following relationship
\begin{eqnarray}
\mathcal{U}^{-1}(\Lambda)P^{\mu}\mathcal{U}(\Lambda)=\Lambda^{\mu}_{\hspace{0.2cm}\nu}\,P^{\nu}\,.
\end{eqnarray}
Therefore, under pure Lorentz boost, eigenvalues of the momentum operator $P^{\mu}$ changes as $p^{\mu}\rightarrow\Lambda^{\mu}_{\hspace{0.2cm}\nu}\,p^{\nu}$. To find out the effect of the Lorentz boosts on the  spin degrees of freedom $\sigma$, it can be assumed that the spin degrees of freedom $\sigma$
form a complete basis, such as any transformation of the state $\vert \mathbf{p},\sigma \rangle$ with a
specific value of $\sigma$ can be represented as a linear superposition of all the possible $\sigma$
states. Therefore, $\mathcal{U}(\Lambda)\vert \mathbf{p},\sigma \rangle$ can be represented as
 \begin{eqnarray}
 \mathcal{U}(\Lambda)\vert p,\sigma \rangle = \displaystyle\sum_{\sigma'}\mathcal{C}_{\sigma'\sigma}(\Lambda,\mathbf{p})\vert \Lambda \mathbf{p},\sigma \rangle\,
 \end{eqnarray}
where the $\mathcal{C}_{\sigma'\sigma}$ are complex numbers and $\Lambda \mathbf{p}$ is the spatial component of the Lorentz transformed four-momentum.\\
To determine the structure of the $\mathcal{C}_{\sigma'\sigma}$ and the effect of the Lorentz boosts $\Lambda$ on a quantum state $\vert p,\sigma \rangle$, we use two invariant quantities, namely
\begin{eqnarray}
p^2\equiv p^{\mu}p_{\mu}=\vec{p}\cdot\vec{p}-E^2&=&-m^2\nonumber\\
p^0&=&E
\end{eqnarray}
Hence, we can use these two invariant quantities
to classify states into specific classes.
For each value of $p^2$ and for each sign ( $p^0$), it is possible
to choose a ‘standard’ 4-momentum $k^{\mu}$ that identifies a specific class of quantum states \cite{Weinberg}. For
massive particles, we can fix the standard 4-momentum $k^{\mu}$ to be the particle’s momentum in the
rest frame, which is $k^{\mu} = (m, 0, 0, 0)$. Then, any 4-momenta $p^{\mu}$ can be expressed in terms of the standard
momentum as 
\begin{eqnarray}
p^{\mu} = (L(p)k)^{\mu} = L^{\mu}_{\hspace{0.2cm}\nu}(p) k^{\nu}
\end{eqnarray} 
where $L(p)$ is a Lorentz transformation that depends on $p^{\mu}$ and takes $k^{\mu} \rightarrow p^{\mu}$. Therefore, quantum states $\vert \mathbf{p},\sigma \rangle$ can be defined in terms of the standard momentum state $\vert \mathbf{k},\sigma \rangle$ as
\begin{eqnarray}
\vert \mathbf{p},\sigma \rangle= N(p)\,\mathcal{U}(L(p))\vert \mathbf{k},\sigma \rangle
\end{eqnarray}
where $N(p)$ is a normalization constant and $\mathbf{k}$ is the spatial component of the standard 4-momentum.\\
Applying a Lorentz boost $\Lambda$ on $\vert \mathbf{p},\sigma \rangle$, we get
\begin{eqnarray}
\mathcal{U}(\Lambda)\vert \mathbf{p},\sigma \rangle&=& N(p)\,\mathcal{U}(\Lambda)\,\mathcal{U}(L(p))\vert \mathbf{k},\sigma \rangle\nonumber\\
&=&N(p)\,\mathcal{U}(I)\,\mathcal{U}(\Lambda L(p))\vert \mathbf{k},\sigma \rangle\nonumber\\
&=&N(p)\,\mathcal{U}(L(\Lambda p))\,\mathcal{U}^{-1}(L(\Lambda p))\,\mathcal{U}(\Lambda L(p))\vert \mathbf{k},\sigma \rangle\nonumber\\
&=&N(p)\,\mathcal{U}(L(\Lambda p))\,\mathcal{U}(L^{-1}(\Lambda p))\,\mathcal{U}(\Lambda L(p))\vert \mathbf{k},\sigma \rangle\nonumber\\
&=&N(p)\,\mathcal{U}(L(\Lambda p))\,\mathcal{U}(L^{-1}(\Lambda p)\Lambda L(p))\vert \mathbf{k},\sigma \rangle\nonumber\\
&=&N(p)\,\mathcal{U}(L(\Lambda p))\,\mathcal{U}(W(\Lambda, p))\vert \mathbf{k},\sigma \rangle
\end{eqnarray}
where $W(\Lambda, \mathbf{p}) = L^{-1}(\Lambda \mathbf{p})\Lambda L(\mathbf{p})$ is called Wigner rotation, which leaves the standard momentum
$\mathbf{k}$ invariant, and only acts on the internal degrees of freedom of $\vert \mathbf{k},\sigma \rangle:\, k\xrightarrow{L} p\xrightarrow {\Lambda}\Lambda p\xrightarrow{L^{-1}} k$. The set of Wigner rotations forms a group known
as the little group, which is a subgroup of the Poincare group. Hence,
the final momentum in the rest frame is different from the original one by a Wigner rotation,
\begin{eqnarray}
\mathcal{U}(W(\Lambda, \mathbf{p}))\vert \mathbf{k},\sigma \rangle=\displaystyle\sum_{\sigma'}\mathcal{D}_{\sigma'\sigma}(W(\Lambda,\mathbf{p}))\vert \mathbf{k},\sigma' \rangle
\end{eqnarray}
where $\mathcal{D}(W(\Lambda,\mathbf{p}))$ is the unitary representation of the little group.\\
On the other hand, $\mathcal{U}(L(\Lambda \mathbf{p}))$ takes $\mathbf{k}\rightarrow\Lambda \mathbf{p}$ without
affecting the spin, by definition. Therefore,
\begin{eqnarray}
\mathcal{U}(\Lambda)\vert \mathbf{p},\sigma \rangle&=& N(p)\,\mathcal{U}(L(\Lambda \mathbf{p}))\,\mathcal{U}(W(\Lambda,\mathbf{p}))\vert \mathbf{k},\sigma \rangle\nonumber\\
&=&N(p)\,\mathcal{U}(L(\Lambda \mathbf{p}))\,\displaystyle\sum_{\sigma'}\mathcal{D}_{\sigma'\sigma}(W(\Lambda,\mathbf{p}))\vert \mathbf{k},\sigma' \rangle\nonumber\\
&=&N(p)\,\displaystyle\sum_{\sigma'}\mathcal{D}_{\sigma'\sigma}(W(\Lambda,\mathbf{p}))\,\mathcal{U}(L(\Lambda \mathbf{p}))\vert \mathbf{k},\sigma' \rangle\nonumber\\
&=&\frac{N(p)}{N(\Lambda p)}\displaystyle\sum_{\sigma'}\mathcal{D}_{\sigma'\sigma}(W(\Lambda,\mathbf{p}))\vert \Lambda \mathbf{p},\sigma' \rangle\,
\end{eqnarray}
where we have used $\vert \Lambda \mathbf{p},\sigma' \rangle= N(\Lambda p)\,\mathcal{U}(L(\Lambda \mathbf{p}))\vert \mathbf{k},\sigma' \rangle$.\\
The ratio $\frac{N(p)}{N(\Lambda p)}$ can be fixed by using appropriate normalization condition.
Hence, under pure Lorentz boosts $\Lambda$, a single particle quantum state $\vert \mathbf{p},\sigma \rangle$ transforms as
\begin{eqnarray}
\mathcal{U}(\Lambda)\vert \mathbf{p},\sigma \rangle=\sqrt{\frac{(\Lambda p)^0}{p^0}}\displaystyle\sum_{\sigma'}\mathcal{D}_{\sigma'\sigma}(W(\Lambda,\mathbf{p}))\vert \Lambda \mathbf{p},\sigma' \rangle\,\label{t-state again}\nonumber\\
\end{eqnarray}
which is eq.(\ref{t-state}) in the main text.


\begin{thebibliography}{8}
\bibitem{AEinsteinPR}
A. Einstein, D. Podolsky, and N. Rosen, ``\textit{Can Quantum-Mechanical Description of Physical Reality Be Considered Complete?}", \href{https://link.aps.org/doi/10.1103/PhysRev.47.777}{Phys. Rev. 47 (1935) 777}.
\bibitem{ESch}
E. Schrodinger, ``\textit{Discussion of Probability Relations between Separated Systems}", \href{https://doi.org/10.1017/S0305004100013554}{Math. Proc. Cambridge Philos. Soc. 31 (2008) 555}.
\bibitem{Bell}
J. S. Bell, ``\textit{On the Einstein Podolsky Rosen paradox}", \href{https://link.aps.org/doi/10.1103/PhysicsPhysiqueFizika.1.195}{Phys. Physique Fiz. 1 (1964) 195}.
\bibitem{Clauser}
J. F. Clauser, M. A. Horne, A. Shimony, and R. A. Holt, ``\textit{Proposed Experiment to Test Local Hidden-Variable Theories}", \href{https://link.aps.org/doi/10.1103/PhysRevLett.23.880}{Phys. Rev. Lett. 23 (1969) 880}; \textit{Erratum:} \href{https://link.aps.org/doi/10.1103/PhysRevLett.24.549}{Phys. Rev. Lett. 24 (549) 1970}.
\bibitem{GingrichAdami}
R. M. Gingrich and C. Adami, ``\textit{Quantum Entanglement of Moving Bodies}", \href{https://link.aps.org/doi/10.1103/PhysRevLett.89.270402}{Phys. Rev. Lett. 89 (2002) 270402}.
\bibitem{Peres}
A. Peres and D. R. Terno, ``\textit{Quantum information and relativity theory}", \href{https://link.aps.org/doi/10.1103/RevModPhys.76.93}{Rev. Mod. Phys. 76 (2004) 93}.
\bibitem{MCzachor}
M. Czachor, ``\textit{Einstein-Podolsky-Rosen-Bohm experiment with relativistic massive particles}", \href{https://link.aps.org/doi/10.1103/PhysRevA.55.72}{Phys. Rev. A 55 (1997) 72}.
\bibitem{Peres2}
A. Peres, P. F. Scudo, and D. R. Terno, ``\textit{Quantum Entropy and Special Relativity}", \href{https://link.aps.org/doi/10.1103/PhysRevLett.88.230402}{Phys. Rev. Lett. 88 (2002) 230402}. 
\bibitem{Alsing}
P. M. Alsing and G. J. Milburn, ``\textit{On entanglement and Lorentz transformations}", \href{https://dl.acm.org/doi/abs/10.5555/2011492.2011496}{Quantum Inf. Comput. 2 (2002) 487}.
\bibitem{Dahn}
D. Ahn, H. J. Lee, Y. H. Moon, and S. W. Hwang, ``\textit{Relativistic entanglement and Bell’s inequality}", \href{https://link.aps.org/doi/10.1103/PhysRevA.67.012103}{Phys. Rev. A 67 (2003) 012103}.
\bibitem{Alsing2}
P. M. Alsing and G. J. Milburn, ``\textit{Teleportation with a Uniformly Accelerated Partner}", \href{https://link.aps.org/doi/10.1103/PhysRevLett.91.180404}{Phys. Rev. Lett. 91 (2003) 180404}.
\bibitem{Peres3}
A. Peres and D. R. Terno, ``\textit{Relativistic doppler effect in quantum communication}", \href{https://www.tandfonline.com/doi/abs/10.1080/09500340308234560}{J. Mod. Opt. 50 (2003) 1165}.
\bibitem{Pachos}
J. Pachos and E. Solano, ``\textit{Generation and degree of entanglement in a relativistic formulation}", \href{http://portal.acm.org/citation.cfm?id=2011520}{Quantum Inf. Comput. 3 (2003) 115}.
\bibitem{Gonera}
C. Gonera, P. Konsi\'{n}ski, and P. Ma\'{s}lanka, ``\textit{Special relativity and reduced spin density matrices}", \href{https://link.aps.org/doi/10.1103/PhysRevA.70.034102}{Phys. Rev. A 70 (2004) 034102}.
\bibitem{Lamata}
L. Lamata, M. A. Martin-Delgado, and E. Solano, ``\textit{Relativity and Lorentz Invariance of Entanglement Distillability}", \href{https://link.aps.org/doi/10.1103/PhysRevLett.97.250502}{Phys. Rev. Lett. 97 (2006) 250502}.
\bibitem{Bartlett}
S. D. Bartlett and D. R. Terno, ``\textit{Relativistically invariant quantum information
}", \href{https://link.aps.org/doi/10.1103/PhysRevA.71.012302}{Phys. Rev. A 71 (2005) 012302}.
\bibitem{TFJordan}
T. F. Jordan, A. Shaji, and E. C. G. Sudarshan, ``\textit{Maps for Lorentz transformations of spin}", \href{https://link.aps.org/doi/10.1103/PhysRevA.73.032104}{Phys. Rev. A 73 (2006) 032104}.
\bibitem{TFJordan2}
T. F. Jordan, A. Shaji, and E. C. G. Sudarshan, ``\textit{Lorentz transformations that entangle spins and entangle momenta}", \href{https://link.aps.org/doi/10.1103/PhysRevA.75.022101}{Phys. Rev. A 75 (2007) 022101}.
\bibitem{Fuentes}
I. Fuentes-Schuller and R. B. Mann, ``\textit{Alice Falls into a Black Hole: Entanglement in Noninertial Frames}", \href{https://link.aps.org/doi/10.1103/PhysRevLett.95.120404}{Phys. Rev. Lett. 95 (2005) 120404}.
\bibitem{Fuentes2}
P. M. Alsing, I. Fuentes-Schuller, R. B. Mann, and T. E. Tessier, ``\textit{Entanglement of Dirac fields in noninertial frames}", \href{https://link.aps.org/doi/10.1103/PhysRevA.74.032326}{Phys. Rev. A 74 (2006) 032326}.
\bibitem{Fuentes3}
G. Adesso, I. Fuentes-Schuller, and M. Ericsson, ``\textit{Continuous-variable entanglement sharing in noninertial frames}", \href{https://link.aps.org/doi/10.1103/PhysRevA.76.062112}{Phys. Rev. A 76 (2007) 062112}.
\bibitem{Lloyd}
S. Lloyd, ``\textit{Almost Certain Escape from Black Holes in Final State Projection Models}", \href{https://link.aps.org/doi/10.1103/PhysRevLett.96.061302}{Phys. Rev. Lett. 96 (2006) 061302}.
\bibitem{ECP}
J. Eisert, M. Cramer, and M. B. Plenio, ``\textit{Colloquium: Area laws for the entanglement entropy}", \href{https://link.aps.org/doi/10.1103/RevModPhys.82.277}{Rev. Mod. Phys. 82 (2010) 277}.
\bibitem{Livine}
E. R. Livine and D. R. Terno, ``\textit{Quantum causal histories in the light of quantum information}", \href{https://link.aps.org/doi/10.1103/PhysRevD.75.084001}{Phys. Rev. D 75 (2007) 084001}.
\bibitem{ASM}
A. S. Majumdar, D. Home, and S. Sinha, ``\textit{Dark energy from quantum wave function collapse of dark matter}", \href{https://www.sciencedirect.com/science/article/pii/S0370269309008594}{Phys. Lett. B 679 (2009) 167}.
\bibitem{Maldacena}
J. Maldacena, ``\textit{A model with cosmological Bell inequalities}", \href{https://doi.org/10.1002/prop.201500097}{Prog. Phys. 64 (2016) 10}. 
\bibitem{TBaumgratz}
T. Baumgratz, M. Cramer, and M. B. Plenio, ``\textit{Quantifying Coherence}", \href{https://link.aps.org/doi/10.1103/PhysRevLett.113.140401}{Phys. Rev. Lett. 113 (2014) 140401}.
\bibitem{Mhorodecki}
M. Horodecki and J. Oppenheim, ``\textit{(QUANTUMNESS IN THE CONTEXT OF) RESOURCE THEORIES}", \href{https://doi.org/10.1142/S0217979213450197}{Int. J. Mod. Phys. B 27 (2013) 1345019}.
\bibitem{Brandao}
F. G. S. L. Brand\~{a}o and G. Gour, ``\textit{Reversible Framework for Quantum Resource Theories}", 
\href{https://link.aps.org/doi/10.1103/PhysRevLett.115.070503}{Phys. Rev. Lett. 115 (2015) 070503}; \textit{Erratum}: \href{https://link.aps.org/doi/10.1103/PhysRevLett.115.199901}{Phys. Rev. Lett. 115 (2015) 070503}.
\bibitem{Gour}
G. Gour and R. W. Spekkens, ``\textit{The resource theory of quantum reference frames: manipulations and monotones}", \href{http://dx.doi.org/10.1088/1367-2630/10/3/033023}{New J. Phys. 10 (2008) 033023}.
\bibitem{Gour2}
G. Gour,I. Marvian, and R. W. Spekkens, ``\textit{Measuring the quality of a quantum reference frame: The relative entropy of frameness}", \href{https://link.aps.org/doi/10.1103/PhysRevA.80.012307}{Phys. Rev. A 80 (2009) 012307}.
\bibitem{Spekkens}
I. Marvian and R. W. Spekkens, ``\textit{The theory of manipulations of pure state asymmetry: I. Basic tools, equivalence classes and single copy transformations}", \href{http://dx.doi.org/10.1088/1367-2630/15/3/033001}{New J. Phys. 15 (2013) 033001}.
\bibitem{Spekkens2}
I. Marvian and R. W. Spekkens, ``\textit{Modes of asymmetry: The application of harmonic analysis to symmetric quantum dynamics and quantum reference frames}", \href{https://link.aps.org/doi/10.1103/PhysRevA.90.062110}{Phys. Rev. A 90 (2014) 062110}.
\bibitem{Spekkens3}
I. Marvian and R. W. Spekkens, ``\textit{Extending Noether’s theorem by quantifying the asymmetry of quantum states}", \href{https://doi.org/10.1038/ncomms4821}{Nat. Commun. 5 (2014) 3821}.
\bibitem{Marvian}
I. Marvian, R. W. Spekkens, and P. Zanardi, ``\textit{Quantum speed limits, coherence, and asymmetry}", \href{https://link.aps.org/doi/10.1103/PhysRevA.93.052331}{Phys. Rev. A 93 (2016) 052331}.
\bibitem{Marvian2}
I. Marvian and R. W. Spekkens, ``\textit{How to quantify coherence: Distinguishing speakable and unspeakable notions}", \href{https://link.aps.org/doi/10.1103/PhysRevA.94.052324}{Phys. Rev. A. 94 (2016) 052324}.
\bibitem{BD}
E. Chitambar and G. Gour, ``\textit{Comparison of incoherent operations and measures of coherence}", \href{https://link.aps.org/doi/10.1103/PhysRevA.94.052336}{Phys. Rev. A 94 (2016) 052336}.
\bibitem{BID}
Y. Yao, G. H. Dong, X. Xiao, and C. P. Sun, ``\textit{Frobenius-norm-based measures of quantum coherence and asymmetry}", \href{https://doi.org/10.1038/srep32010}{Sci. Rep. 6 (2016) 32010}.
\bibitem{BID2}
W.-C. Wang, M.-F. Fang, and M. Yu, ``\textit{Intrinsic basis-independent quantum coherence measure}", \href{https://doi.org/10.48550/arXiv.1701.05110}{arXiv: 1701.05110}.
\bibitem{RiddhiASM}
R. Chatterjee and A. S. Majumdar, ``\textit{Preservation of quantum coherence under Lorentz boost for narrow uncertainty wave packets}", \href{https://link.aps.org/doi/10.1103/PhysRevA.96.052301}{Phys. Rev. A 96 (2017) 052301}.
\bibitem{Weinberg}
S. Weinberg, \href{https://www.cambridge.org/core/books/quantum-theory-of-fields/22986119910BF6A2EFE42684801A3BDF}{The Quantum Theory of Fields: Volume I: Foundations}, Cambridge University Press (1995), Cambridge.
\end{thebibliography}
\end{document}